\documentclass[aps,pre,amsmath,amsfonts,amssymb,floatfix,superscriptaddress,nofootinbib]{revtex4}
\usepackage{mathrsfs} \usepackage{graphicx,color} \usepackage{bbm} \usepackage{multirow}
\usepackage{algorithm,algpseudocode}

\usepackage{rotate}
\usepackage{epsfig}
\usepackage{psfrag}
\usepackage{mathtools}

 \let\b=\beta \let\g=\gamma \let\d=\delta
  \let\h=\eta

   \let\G=\Gamma
 \let\Th=\Theta  

\let\r=\rho  \let\io=\infty

\def\PP{{\cal P}}\def\EE{{\cal E}}

\def\LL{{\cal L}}  
\def\DD{{\cal D}} 
  \def\XX{{\cal X}}

\def\to{\rightarrow}

\newcommand{\beq}{\begin{equation}} \newcommand{\eeq}{\end{equation}}

\newcommand{\avg}[1]{\langle #1\rangle}

\newcommand\be{\begin{equation}}
\newcommand\bea{\begin{eqnarray} \nonumber }
\newcommand\ee{\end{equation}}
\newcommand\eea{\end{eqnarray}}

\begin{document}

\title{Monetary Policy and Dark Corners in a stylized Agent-Based Model}

\author{Stanislao Gualdi}
\affiliation{
Laboratoire de Math\'ematiques Appliqu\'ees aux Syst\`emes, \'Ecole Centrale Paris, 92290 Ch\^atenay-Malabry, France
}
\email{stanislao.gualdi@gmail.com}

\author{Marco Tarzia}
\affiliation{
Universit\'e Pierre et Marie Curie - Paris 6, Laboratoire de Physique Th\'eorique de la Mati\`ere 
Condens\'ee, 4, Place Jussieu, 
Tour 12, 75252 Paris Cedex 05, France }

\author{Francesco Zamponi}
\affiliation{Laboratoire de Physique Th\'eorique,
\'Ecole Normale Sup\'erieure, UMR 8549 CNRS, 24 Rue Lhomond, 75231 Paris Cedex 05, France}

\author{Jean-Philippe Bouchaud}
\affiliation{CFM, 23 rue de l'Universit\'e, 75007 Paris, France, and Ecole Polytechnique, 91120 Palaiseau, France.}

\begin{abstract} 
We extend in a minimal way the stylized macroeconomic Agent-Based model introduced in our previous paper~\cite{Tipping}, with the 
aim of investigating the role and efficacy of monetary policy of a `Central Bank' that sets the interest rate such as 
to steer the economy towards a prescribed inflation and employment level. 
Our major finding is that provided its policy is not too aggressive (in a sense detailed in the paper) 
the Central Bank is successful in achieving its goals. 
However, the existence of different equilibrium states of the economy, separated by phase boundaries (or ``dark corners''), 
can cause the monetary policy itself to trigger instabilities 
and be counter-productive. In other words, the Central Bank must navigate in a narrow window: too little is not enough, too much 
leads to instabilities and wildly oscillating economies. This conclusion strongly contrasts with the prediction of DSGE models. 
\end{abstract} 

\maketitle

\clearpage
\tableofcontents

\clearpage

\section{Introduction}

{
The Agent Based Model (ABM) studied in this paper is a {generalization} of the ``toy-ABM'' (dubbed Mark-0) recently introduced in~\cite{Tipping}, 
following previous work by the group of Delli Gatti et al.~\cite{DelliGatti}. 
}
Mark-0 considers a {stylized} economy with firms and households, but no banks, 
no interest rates on loans and deposits, and therefore no direct concept of ``monetary policy''. As discussed at length in~\cite{Tipping}, 
the original motivation of Mark-0 was mostly to illustrate the importance of phase diagrams and phase transitions in the context of ABMs, 
in particular the sensitivity of the state of the (artificial) economy on a subset of parameters. 
Small changes in the value of these parameters were indeed found to induce sharp variations in aggregate output, 
unemployment or inflation. In other words, endogenous crises can occur in such economies, as the result of insignificant or anecdotal
changes in the environment. This possibility is quite interesting in itself, and must be contrasted with more traditional economic models, 
such as the popular DSGE framework~\cite{DSGE,Review}, where the dynamics is linear and only large exogenous shocks can cause havoc. 
As recently pointed out by O. Blanchard in a very inspiring piece~\cite{Blanchard}: {\it 
We in the field did think of the economy as roughly linear, constantly subject to different shocks, constantly fluctuating, 
but naturally returning to its steady state over time. [...]. The main lesson of the crisis is that we were much closer to ``dark corners'' -- 
situations in which the economy could badly malfunction -- than we thought.}

Because they can deal with non-linearities, heterogeneities and crises, ABMs are often promoted as possible alternatives to the DSGE models 
used in central banks as guides for monetary policy~\cite{Eurace,SantAnna,Lagom,Review}. It is therefore clear that introducing interest rates 
monitored by a central bank in Mark-0 is mandatory for policy makers to develop any interest in the ABM research program in general and our model in 
particular. The aim of the present paper is to 
parsimoniously extend Mark-0 as to capture the effects of monetary policy on the course of the economy. We first identify and model several channels through which 
interest rates can feed into the {behavior} of firms and households. We then study different policy experiments, whereby the ``Central Bank'' attempts 
to reach a target inflation and/or unemployment level using a Taylor rule to set the interest rate (see Eq. (\ref{taylor_rule}) below). We find that 
provided the economy is far from phase boundaries (or ``dark corners''~\cite{Blanchard}) such policies can be successful, whereas too aggressive policies 
may in fact, unwillingly, drive the economy to an unstable state, where large swings of inflation and unemployment occur. 

{
Our Agent-Based framework is voluntarily bare bones. It posits a minimal set of plausible ingredients that are most probably present in the real world in one form
or the other. For example, we assume reasonable heuristic rules for the hiring/firing and wage policies of firms confronted with over- or under-production, 
or with a rising level of debt. Similarly, our model encodes in a schematic manner the consumption {behavior} of households facing inflation
and rising rates, that is in fact similar to the standard Euler equation for consumption in general equilibrium/DSGE models~\cite{DSGE}. Our 
approach is therefore prone to the usual critique addressed to ABMs: the rules we implement are -- although reasonable -- to some degree arbitrary.
The ABM community is of course aware of this weakness, with many attempts to resolve it, such as imposing consistency constraints on behavioral 
assumptions (see~\cite{Dawid}, Appendix A), or, even better, relying on micro-panel data that reveal how firms and households actually make decisions under 
different macro-economic or specific conditions (see for example~\cite{Lein,rates} for the behavior of firms, and~\cite{Souleles,Ludvigson,recession} for the behaviour
of households). However, reliable empirical data are still rather scarce and do not allow yet to answer all the questions needed to constrain and 
calibrate an ABM, even simple ones like Mark-0. 
}

{
Our philosophy, explained in detail in~\cite{Tipping}, is different. We argue that the {\it qualitative}, aggregate {behavior} elicited by the Mark-0 model is in fact robust and generic, 
although the actual {\it quantitative} aspects may not be (as, for example, the precise value of the parameters of the Taylor rule beyond which instabilities occur). In other 
words, if one forgoes  the idea of quantitatively predicting the macro-behaviour of the economy but is satisfied (at least temporarily) with a qualitative description
of the possible aggregate behaviour, some progress is possible without detailed knowledge of the micro-rules. Our belief is backed by the idea --  pervasive in many areas of science -- 
that the aggregate properties of interacting entities can be classified in different phases, separated by phase boundaries across which radical changes of the 
emergent {behavior} take place. This idea has a long history, in particular in economics -- remember, for example, the title of Thomas Schelling's famous book:
{\it Micromotives and Macrobehaviour}~\cite{Schelling}; for more recent progress see e.g.~\cite{Kirman,Hommes,Marsili,Marsili2,Haldane,Joao} and 
for recent reviews:~\cite{JPB,Marsili-review}. To bolster our belief that emergent aggregate properties are robust against micro-changes, we have tested many variants 
of the model presented below and indeed found that the overall {behavior} of our artificial economy is remarkably robust -- in particular the presence of instabilities 
and crises. Following up on O. Blanchard's lament~\cite{Blanchard}, the existence of large swaths of the parameter space where the economy is prone to violent crises 
seems to be an unavoidable fact that we have to learn to confront with~\cite{Kirman,Caballero}. 
}

{
The work presented in this paper is part of a growing literature of macroeconomic agent-based model~\cite{testfatsion1,testfatsion2}.
In the recent years several such models have been developed, allowing both to 
reproduce macro-stylized facts and to study policy design~\cite{Dawid-review}. Building upon~\cite{SantAnna}, Dosi et al.~\cite{Dosi1,Dosi2} characterize the 
impact of fiscal and monetary policies on macroeconomics fluctuations. 
Fiscal policies are found to have a greater role in dampening business cycles, reducing unemployment 
and the likelihood of economic crises. Another example somewhat related to our work is Ref.~\cite{Lengnick1}, 
which focuses on a simple ABM of financial markets 
coupled with a New Keynesian DSGE model, which allows one to  
reproduce endogenously stock price bubbles and business cycles and to study the 
introduction of financial taxes. Calibration and validation procedures have  
been explored in~\cite{Haber,Bianchi,Fagiolo}.
}

{
Although similar in spirit, the present study builds upon a slightly different perspective,
following the framework and philosophy developed in~\cite{Tipping}.
In particular, we model an idealized closed economy with linear production capabilities, 
no capital (labor is the only input for production), no innovation and growth, 
no financial sector, and only a minimal set of additional behavioural rules which couple the Central Bank
monetary policy to firms' and agents' choices. In this sense, our model is much more parsimonious than what is found in
the recent ABM literature and voluntarily overlooks important actors and processes at play in the real world.
Our central concern is instead to characterize the qualitative behaviour of the emerging economy and 
investigate its dynamical stability  with respect to the Central Bank's policy. 
}

{
The outline of the paper is as follows. In Section II we give a brief description of the Mark-0 model proposed in~\cite{Tipping} and 
detail the minimal additional rules we need to couple the monetary policy with firms' and agents' decisions. In Section III, we investigate
the ``natural'' state of our toy-economy, in the presence of interest rates but without any Central Bank intervention. In Section IV, we perform
several policy experiments: an expansionary monetary shock, and the implementation of a Taylor-rule monitoring of the interest rate by 
the Central Bank to achieve given employement and inflation targets. We establish the phase diagram of our model in the presence of this 
Taylor-rule based intervention. We briefly compare our findings to the prediction of DSGE models. We conclude in Section V. Appendix A 
gives further details about Mark-0; Appendix B investigates the continuous phase transition induced by the wary behaviour of indebted firms; 
finally, Appendix C gives a full pseudo-code of our model that should allow easy duplication of our results.
}

\section{Description of the model}

\subsection{Brief summary of the minimal ``Mark-0'' model}

The Mark-0 closed economy is made of firms and households. While the latter sector is represented at an aggregate
level, firms are heterogeneous and treated individually. Each firm $i=1, \dots, N_F$ at time $t$ produces a quantity $Y_i(t)$ of perishable goods
that it attempts to sell at price $p_i(t)$. It needs a number of $N_i(t)=Y_i(t)/\zeta_i$ of employees\footnote{$\zeta_i$ is the 
productivity of firm $i$. We chose $\zeta_i = 1$ in~\cite{Tipping} and we will stick to this choice throughout the 
present paper as well.} to produce $Y_i(t)$, and pays  a wage $W_i(t)$. 
The demand $D_i$ for good $i$ depends on the global consumption budget of households $C_B(t)$, itself determined as 
a fraction of the household savings (that include the last wages), and it is a decreasing function of the asked price $p_i(t)$, with a price sensitivity 
parameter that can be tuned -- see Appendix A.

To update their production, price and wage policy, firms use reasonable ``rules of thumb''~\cite{Tipping} that we detail 
in Appendix A and that were already (partly) justified in the original work of Delli Gatti et al.~\cite{DelliGatti}.
For example, production is decreased and employees are made redundant whenever $Y_i > D_i$, and vice-versa.\footnote{As a consequence 
of these adaptive adjustments, 
the economy can reach (in some regions of the parameter space) equilibrium, corresponding to the market clearing condition one would obtain 
in a fully representative agent framework. However, fluctuations around equilibrium persists in the limit of large 
system sizes giving rise to a rich phenomenology, see~\cite{DelliGatti,Tipping}.}
The adjustment speed can however be asymmetric, i.e. the ratio $R$ of hiring adjustement speed to firing adjustement speed is not necessarily equal to one, for
example because of labour laws. 
This turns out to be one of the
most important control parameter that determines the fate of the overall economy.

In the initial state of the economy, firms are heterogenous in size and prices (with a uniform distribution around the average 
size and price, see Appendix C for details), but all offer the same wage and see the same demand. This is arbitrary and of 
little importance, because the stationary state of the model is statistically independent of the initialization.

When the Mark-0 economy is set in motion, it soon becomes clear that some firms have to take up loans in order to stay in business. 
One therefore immediately has to add further rules for this to take place. In the zero-interest rate world of Mark-0, we let firms freely 
accumulate a total debt up to a threshold that is a multiple $\Theta$ of total payroll, beyond which the firm is declared bankrupt 
(its debt is then repaid partly by households and partly by surviving firms, such that there is no net creation of money). 
From this point of view the parameter $\Theta$ determines the maximum credit supply available to firms. Fixing the value of $\Theta$ 
plays the role of a primitive monetary policy, since the total amount of money circulating in the economy (`broad money') directly depends on $\Theta$~\cite{Tipping}. 
When $\Theta=0$, no debt is allowed (zero leverage), while when $\Theta \to \infty$, firms have not limit on the loans they need to continue business 
(unbounded leverage). 

\begin{figure}
\centering
\includegraphics[scale=0.35]{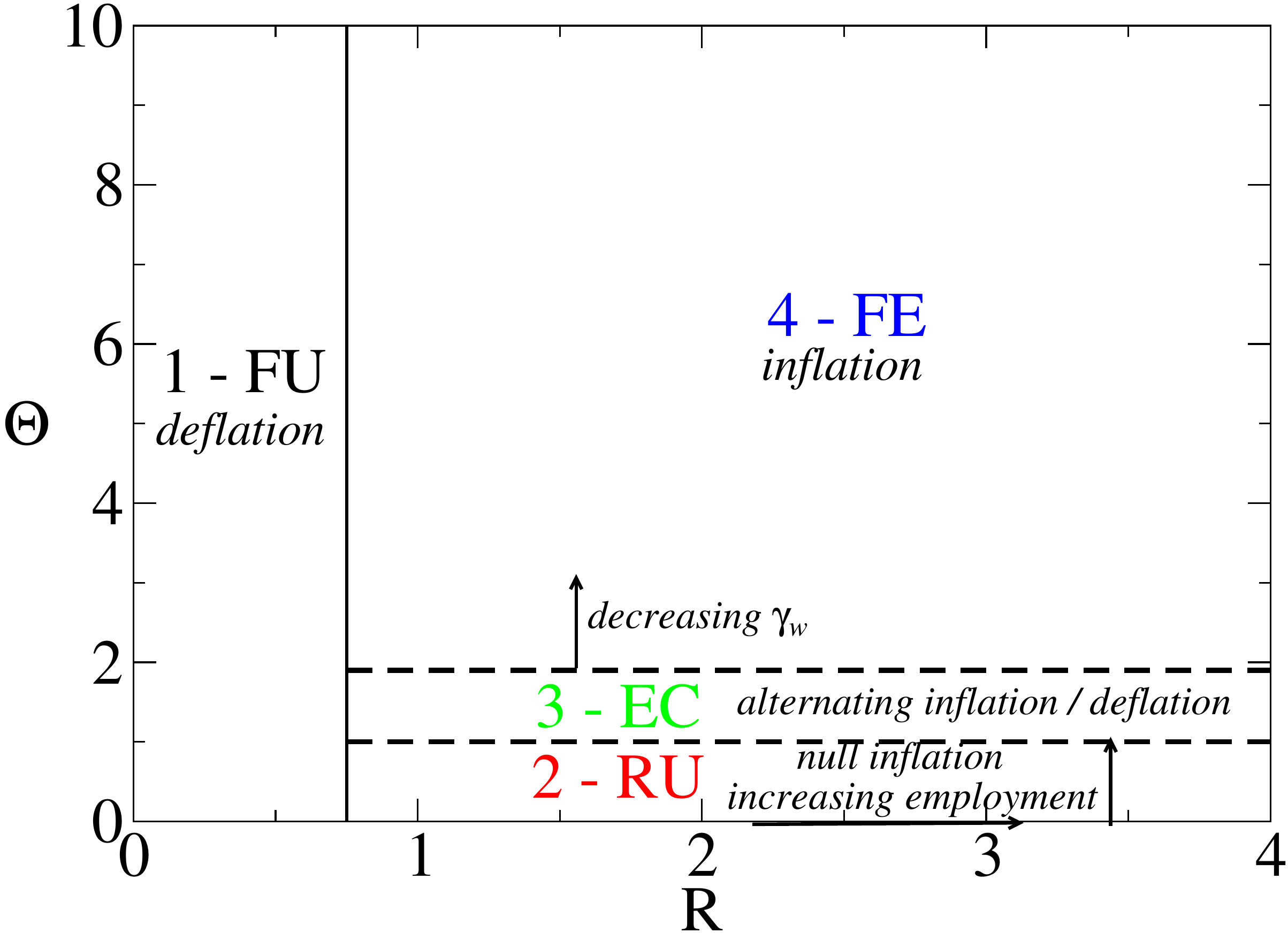}
\includegraphics[scale=0.35]{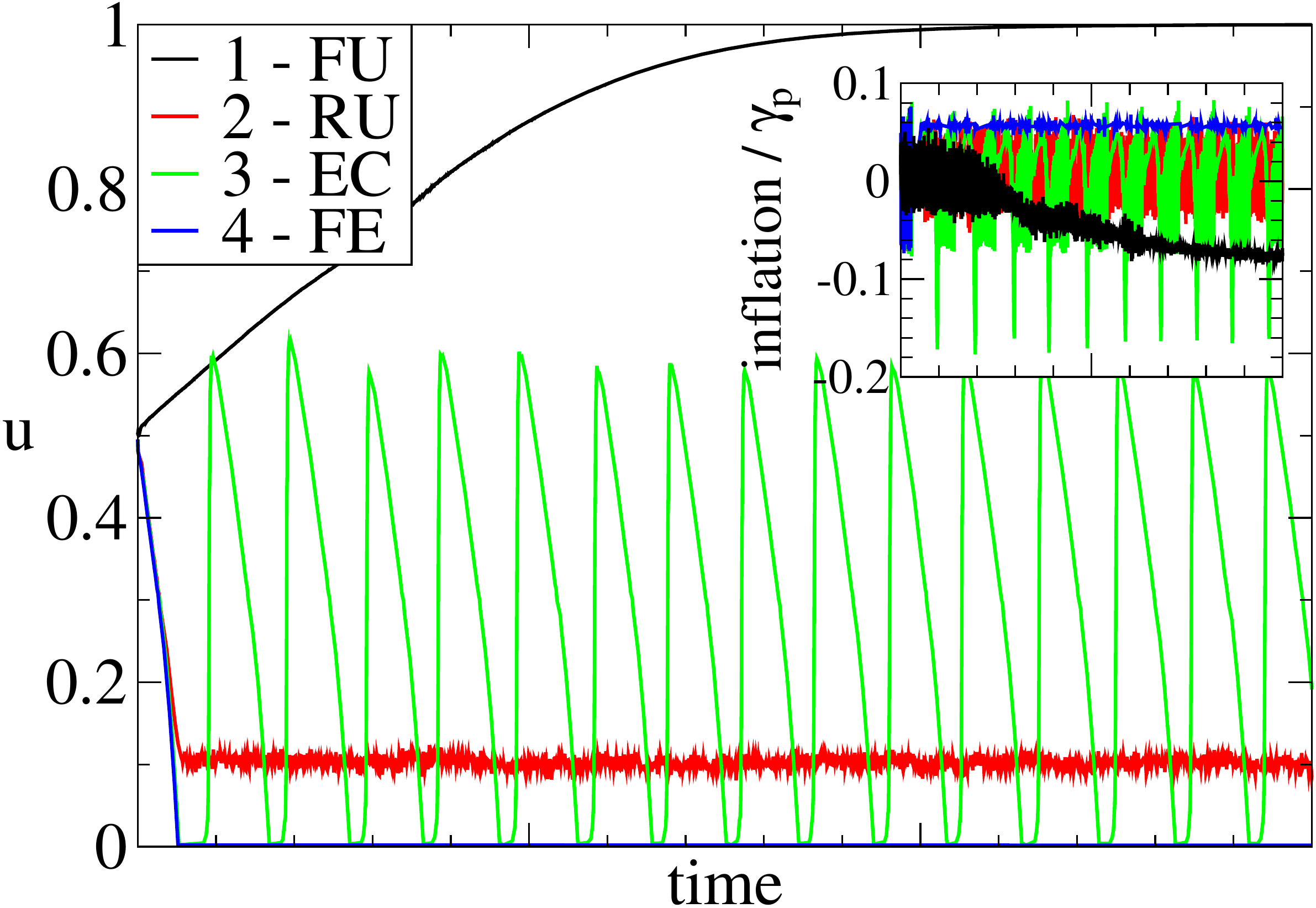}
\caption{
({\it Left})
Phase diagram in the $R-\Theta$ plane of the basic Mark-0 model as obtained
in~\cite{Tipping} with wage update. There are four distinct phases separated by critical 
lines. The Full Employment (FE) phase ($R > R_c$, $\Theta$ large) is {characterized} by positive average inflation, 
while there is deflation in the Full Unemployment (FU) phase ($R < R_c$). Endogenous Crises (EC) are {characterized} by alternating 
cycles of inflation and deflation, and occur for $R > R_c$, $\Theta$ intermediate. Finally, $R > R_c$, small $\Theta$ correspond to
a region of small inflation and Residual Unemployment (RU).
The location of phase boundaries is only weakly affected by the choice of the other parameters of 
Mark-0, see~\cite{Tipping}.
({\it Right})
Typical trajectories of the unemployment rate $u(t)$ for each of the phases. In the inset, the price 
variations are shown, displaying either inflation (in the FE phase) or deflation (in the FU phase). $\gamma_p^{-1}$ (resp. $\gamma_w^{-1}$) sets the characteristic time scale
for price adjustments (resp. wages) in the model, see Appendix A. The surprising occurrence of endogenous oscillations in the EC phase can be fully understood analytically, see~\cite{Synchro}.
}
\label{fig:PD_wages}
\end{figure}

While there are several other parameters needed to define completely Mark-0 (9 parameters in total, see Appendix A), the detailed investigation of~\cite{Tipping} has established 
that only $R$ and $\Theta$ are the ones determining the phase-diagram of the model, shown in
Fig.~\ref{fig:PD_wages}, where we also plot typical trajectories
of the economy in each phase. It is important to stress that this diagram is extremely robust against both details of
the model specification and the value of the other parameters, which only affect the above phenomenology quantitatively, 
but leave the qualitative emergent {behavior} essentially unchanged. Its salient features are~\cite{Tipping}:

\begin{itemize}
 \item When $\Theta=\infty$ the economy is {characterized} by two distinct 
 phases separated by a first order (discontinuous) phase transition as a function of the parameter $R$. When $R<R_c$ (i.e. fast downward production adjustments), 
 one finds at long times a collapse of the economy towards a deflationary/low demand/full unemployment state (FU). 
 For $R>R_c$, on the other hand, the long run state of the
 economy is characterized by a positive inflation/high demand/full employment phase (FE). 
 \item When $\Theta<\infty$ the above description holds but must be refined to allow for the appearance of three sub-phases for $R>R_c$:
 \begin{enumerate}
 \item a full employment and inflationary phase for high values of $\Theta$ (the FE phase, similar to the $\Theta=\infty$ case);
 \item a phase for intermediate values of $\Theta$ {characterized} by high employment and inflation on average, {which is, however,} intermittently disrupted by ``endogenous crises'' (EC), 
 that temporarily bring deflation and high unemployment spikes;
 \item a phase with zero inflation and residual unemployment for small $\Theta$ (the RU phase), where the impossibility to obtain loans creates a positive 
 stationary level of bankruptcies. 
\end{enumerate}
\end{itemize}

Note that the unexpected occurrence of purely endogenous oscillations in the EC phase can in fact be understood fully 
analytically~\cite{Synchro} and {it} is a robust feature that only relies 
on very mild assumptions about the destabilizing feedback mechanisms present in the economy.
 
\subsection{Introducing interest rates in Mark-0}
\label{m0extended}

We now introduce in the model a banking system made up of a Central Bank (CB) which sets the base interest rate $\rho_0(t)$ (and, as 
part of a prudential policy, the parameter $\Theta$ that controls the maximum credit supply available to firms), and a private 
banking system that will act as a transmission belt for the CB policy, by setting interest rates on deposits 
($\rho^{\text{d}}(t)$) and on loans ($\rho^{\ell}(t)$). These interest rates will in turn impact the economy through three channels that 
we detail below: a) direct cost of loans and gains on deposits; b) {behavior} of the firms; c) {behavior} of households.

\subsubsection{The Central Bank policy}

The CB attempts to steer the economy towards a target inflation level $\pi^*$ and employment level $\varepsilon^*$ (equivalent to 
a target output, since the productivity $\zeta$ is set to unity in the present version of the model). The instantaneous inflation $\pi(t)$ and employment
level $\varepsilon(t)$ are defined here as:
\be
\pi(t) = \frac{\overline p(t) - \overline p(t-1)}{\overline p(t-1)}; \qquad \overline p(t) = \frac{\sum_i p_i(t) Y_i(t)}{\sum_i Y_i(t)},
\ee
where $\overline p(t)$ is the production-weighted average price, and:
\be
\varepsilon(t) = \frac{1}{N} \sum_i Y_i(t) \ , \qquad u(t) = 1 - \varepsilon(t),
\ee
where $N$ is the total workforce, and $\varepsilon(t),u(t)$ are respectively the employment and unemployment rate. 

We will assume that the monetary policy followed by the CB for fixing the base interest rate is described by a standard 
Taylor rule of the form~\cite{Mankiw,DSGE}:\footnote{In Gal\`\i's reference book~\cite{DSGE}, the quantity $\phi_\varepsilon$ 
is noted $\phi_y$.}
\be
\label{taylor_rule}
\rho_0(t) = \max\left(\rho^* + 10\ \phi_\pi [\widetilde{\pi}(t)-{\pi}^*] + 
\phi_\varepsilon \log{[\widetilde{\varepsilon}(t)/\widehat{\varepsilon}^*]},\, 0\right)
\ee
where $\rho^*$ is the ``natural'' interest rate that would prevail if the target inflation ${\pi}^*$ and 
target employment $\varepsilon^*$ were reached, and $\alpha_{\pi,\varepsilon}>0$ quantify the intensity of the policy
(high values of the parameters correspond to aggressive policies). Note that $\rho_0$ is constrained to be non-negative. 
The factor $10$ in front of $\phi_\pi$ is only there for convenience, such
that interesting values of $\phi_\pi$ and $\phi_\varepsilon$ are of the same order of magnitude. The notation $\widetilde{x}(t)$ 
corresponds to the exponential moving average of the variable $x(t)$, 
defined as:\footnote{We chose $\omega=0.2$, which corresponds to an averaging time of
about $-1/\log{(1-\omega_\tau)}\approx 4.5$ time steps.} 
\be
\widetilde{x}(t+1) = \omega x(t) + (1-\omega)\widetilde{x}(t).
\ee
In order to avoid unnecessary excessive policy response when the target employment rate is too far from the actual employment rate, 
we actually define a one-time-step target employment rate $\widehat{\varepsilon}^*$ as
\be
\widehat{\varepsilon}^*=\min{\{1.025\ \widetilde{\varepsilon}(t),{\varepsilon}^*\}}
\ee 
meaning that if the employment rate is much lower than the policy target ${\varepsilon}^*$ the CB will try to 
increase it by $2.5 \%$ at each time step until the target is reached. Such a regularization was found not to be needed for inflation.

\subsubsection{The banking sector}

Households in the Mark-0 economy cannot borrow and are thus {characterized} by their total savings $S(t) \geq 0$. Firms, on the other 
hand, can have either deposits ($\EE_i>0$) or liabilities ($\EE_i<0$). Defining $\EE^+=\sum_i\max{(\EE_i,0)}$ and $\EE^-= -\sum_i\min{(\EE_i,0)}$, 
the balance sheet of the banking system reads:
\be
M + \EE^-(t) = S(t) + \EE^+(t) \equiv \mathcal{X}(t),
\ee
where $M$ is the amount of currency (or initial deposits) created by the central 
bank, which is kept fixed in time, and $\mathcal{X}$ is the total amount of deposits, 
therefore to initial deposits $M$ plus the money created by the banking system when 
issuing loans. 

We now assume that the banking sector fixes the interest rates on deposits and loans ($\rho^{\text{d}}(t)$ and $\rho^{\ell}(t)$ respectively)
uniformly for all lenders and borrowers according to the following rules:\footnote{Note that the parameter $f$ is similar to, but different from the
parameter also called $f$ in~\cite{Tipping}, which was used to share the cost of defaults on firms and households.}
\bea
 \label{interests}
 \rho^{\ell}(t) &=& \rho_0(t) + f\frac{\DD(t)}{\EE^-(t)}  \\
 \rho^{\text{d}}(t) &=& \frac{\rho_0(t)\EE^-(t)-(1-f) \DD(t)}{\XX(t)} \ .
\eea
where $\DD(t)$ is the aggregate costs coming from all firms that just defaulted, bearing on the banking sector. Note that $\rho^{\text{d}} \leq \rho_0 \leq \rho^{\ell}$. 
The parameter $f \in [0,1]$  
reflects the impact of these defaults -- either entirely on the cost of loans ($f=1$) or on the revenue of deposits ($f=0$).
The logic behind rule Eq. (\ref{interests}) is that the interest rate on loan increases when defaults increase, while the rate on deposits is chosen in such a way that 
the profits of the banking sector are exactly zero at each time step. Indeed, one has, for any value of $f$:
\be
\underbrace{\rho^{\ell}(t) \EE^-(t)}_{\text{interest from loans}} \quad - 
\underbrace{\rho^{\text{d}}(t) \XX(t)}_{\text{interest paid on deposits}} \quad -  
\underbrace{\DD(t)}_{\text{cost of defaults}} 
\equiv \, 0.
\ee
Note that $\rho^{\text{d}}(t)$ can become negative for large enough $\DD(t)$ when $f \neq 1$, i.e. the deposits are taxed by the banking sector.
This could be interpreted as a kind of ``bail-in tax'' to absorb debt
in extreme cases. In our simulations where $f=0.5$, this situation does occur in the unstable phases of the economy {during a relatively short fraction of the
time, corresponding to the peaks of the unemployment spikes}.  
Finally, one could introduce an extra haircut in $\rho^{\ell}$ or $\rho^{\text{d}}$
if one wants to model a profit-seeking banking sector, but the resulting profits would somehow have to be re-injected in the economy -- an extra {modeling}
step that we avoid at this stage by assuming the above no-profit rule.

\subsubsection{Households' consumption budget}

As mentioned above, one major simplification of Mark-0 is to treat the whole household sector at the aggregate level, and {to represent it with}
only a few variables: total savings $S(t)$, total wages $W_T(t) = \sum_i W_i(t)Y_i(t)$, and total consumption budget $C_B(t)$ (which, as emphasized 
in~\cite{Tipping}, is in general larger than the actual consumption $C(t)$).

The effect of interest rates on households is two-fold. First, quite trivially, they receive some interest on their savings $S(t)$ that adds to 
the wages $W_T(t)$ and dividends as their total income. Second, the comparison between interest rates and 
inflation creates an incentive to consume or to save. This is in the spirit of the standard Euler equation of DSGE models where consumption is 
found to depend on the difference of rates on deposits $\rho^{\text{d}}(t)$ and inflation $\pi(t)$ (see e.g.~\cite{Mankiw,DSGE}). 
We therefore posit that the consumption budget of households $C_B(t)$ is given by:
\be
\label{cons_budget}
C_B(t) = c(t) \left[ S(t) + W_T(t) + \rho^{\text{d}}(t)S(t) \right]\quad \text{with} \quad c(t)=c_0 \left[1+\alpha_c (\widetilde{\pi}_t-\widetilde{\rho}^{\text{d}}_t)\right],
\ee
where $c(t)\in[0,1]$ is the consumption propensity (taken to be a constant in Mark-0) and $\alpha_c>0$ is a coupling constant that determines the sensitivity 
of households to the (moving average of the) difference between inflation and the interest paid on their savings. The larger the difference between the two, the larger the
propensity to consume, as in standard equilibrium models~\cite{Mankiw}, but with here undetermined phenomenological parameters $c_0,\alpha_c$ that should in principle be measured 
on micro-data (surveys, laboratory experiments, etc., see e.g.~\cite{Souleles,Ludvigson,recession}). 

Apart from these changes, the {behavior} of households is exactly the same as in Mark-0, see~\cite{Tipping} and Appendix A for details.  

\subsubsection{Firms' policy when debt is costly}

\paragraph{Financial fragility.} 

Unlike households the $N_{\rm F}$ firms are heterogeneous and treated individually. 
Each firm is characterized by its production $Y_i$ (equal to its workforce), demand for its goods $D_i$, price $p_i$, wage $W_i$ and cash balance $\EE_i$.
The debt level of a firm is measured through the ratio
\be
\Phi_i=-\EE_i/(W_i Y_i),
\ee
which we interpret as an index of financial fragility.
If $\Phi_i(t) < \Theta$, i.e. when the flux of credit needed from the bank is not too high compared 
to the size of the company (measured as the total payroll), the firm is allowed to continue
its activity. If on the other hand $\Phi_i(t) \geq \Theta$, the firm
defaults and contributes to total default costs $\DD(t)$.

\paragraph{Production and wage update.} 

If the firm is allowed to continue its business, it adapts its price, wages and productions according to
reasonable ``rules of thumb'' introduced in~\cite{Tipping} -- see Appendix A. In particular, the production update is chosen as:
 \beq
 \label{y_update}
\begin{split}
    \text{If } Y_i(t) < D_i(t)  &\hskip10pt \Rightarrow \hskip10pt 
     Y_i(t+1)=Y_i(t)+ \min\{ \eta^+_i ( D_i(t)-Y_i(t)), u^*_i(t) \} \\
    \text{If }   Y_i(t) > D_i(t)  &\hskip10pt  \Rightarrow \hskip10pt
    Y_i(t+1)= Y_i(t) - \eta^-_i [Y_i(t)-D_i(t)]  \\
\end{split}
\eeq
where $u^*_i(t)$ is the maximum number of unemployed workers available to the firm $i$ at time $t$ (see Appendix A). 
The coefficients $\eta^\pm \in [0,1]$ express the sensitivity of the firm's target production to excess 
demand/supply, and they were constant in Mark-0. 

Here, we further postulate that the production adjustment depends on the financial fragility $\Phi_i$ of the firm.
Firms that are close to bankruptcy are arguably faster to fire and slower to hire, and vice-versa for healthy firms. In order to
model this tendency, we posit that the coefficients $\eta^\pm_i$ for firm $i$ are given by: 
\bea
\eta^-_i = \eta_0 \max(1+ \Gamma \Phi_i(t),0) \\
\eta^+_i = R \eta_0 \max(1 - \Gamma \Phi_i(t),0) 
\eea 
where $\eta_0$ is a fixed coefficient, identical for all firms, and $R$ is the propensity ratio discussed 
in the previous section. The factor $\Gamma > 0$ measures how the financial fragility of firms influences their 
hiring/firing policy, since a larger value of $\Phi_i$ then leads to a faster downward adjustment of the workforce 
when the firm is over-producing, and a slower (more cautious) upward adjustment when the firm is under-producing.
The above definition however ensures that $\eta^\pm$ always remain non-negative, i.e. the reaction of the firms is 
always in the intuitive direction.

It is plausible that the financial fragility of the firm also affects its wage policy: we give in Appendix A the 
wage update rules of Mark-0 and their modification to account for financial fragility, through the very same 
parameter $\Gamma$. In essence, deeply indebted firms seek to reduce wages more rapidly, whereas flourishing firms tend to
increase wages quickly. 

The baseline Mark-0 model corresponds to $\Gamma \equiv 0$, and leads to the phase diagram shown in Fig.~\ref{fig:PD_wages} above.
Interestingly, a non-zero value of $\Gamma=\Gamma_0$ (constant across firms and in time) changes substantially the 
{\it nature} -- but not the existence -- of the phase transition between the full employment (FE) 
and full unemployment (FU) phase.  The first order (discontinuous) 
transition for $\Gamma_0=0$, $\Theta=\infty$ found in~\cite{Tipping} and shown in Fig.~\ref{fig:PD_wages}, 
is replaced by a second order (continuous) transition when the firms adapt their {behavior}
as a function of their financial fragility, i.e. when $\Gamma_0 > 0$. Moreover, the ``bad'' FU phase for $R < R_c$ becomes 
a partial unemployment phase with $u < 1$ that continuously varies with $R$: see Appendix B and Fig.~\ref{fig:EmpRes} for full details. 
As firms become more careful, employment can be to some extent preserved -- as expected.

The ``good'' phase of the economy, on the other hand, is only mildly affected by a non zero $\Gamma_0$ -- for example the FE region of Fig.~\ref{fig:PD_wages} 
expands downwards, which is expected since firms manage more carefully their balance sheet, reducing the occurrence of defaults.

\paragraph{The influence of interest rates on the strategy of firms.} 

We now argue that $\Gamma$ should in fact depend on the difference between the interest rate and the inflation: 
high cost of credit makes firms particularly wary of going into debt and their sensitivity to their financial 
fragility should be increased. Therefore, we postulate that interest rates feedback into the {behavior} of the firm 
primarily through the $\Gamma$ parameter, that we model in the simplest possible way as:
\be
\label{alpha_Gamma}
\Gamma = \max{\{\alpha_\Gamma (\widetilde\rho^{\ell}(t)-\widetilde{\pi}(t)),\Gamma_0\}},
\ee
where $\alpha_\Gamma$ (similarly to $\alpha_c$ above) captures the influence of the real interest rate $\widetilde\rho^{\ell}(t)-\widetilde{\pi}(t)$ 
on the hiring/firing policy of the firms. Whenever the real interest rate stays below $\Gamma_0/\alpha_\Gamma$, 
the response of firms to changes of interest rates is negligible (perhaps as reported in~\cite{rates}), whereas larger rates lead to a substantial
change in the firms policy. The case $\alpha_\Gamma = 0$ but $\Gamma_0 > 0$ corresponds to the above discussion and 
{it} is interesting in itself (see Appendix B). However, since we will be mostly concerned with policy issues, we will
concentrate below on the other extreme case, $\alpha_\Gamma > 0$ and $\Gamma_0 = 0$, keeping in mind that reality is 
probably in-between. Note that $\Gamma$ as defined above is now zero when real interest rates are negative, and is 
positive otherwise. 

\subsection{Summary}

\subsubsection{How many new assumptions?}

{
In our attempt to include interest rates and monetary policy into the Mark-0 framework, we have made new behavioral assumptions, that we tried 
to keep as simple and as parsimonious as possible:
\begin{itemize}
\item For the Central Bank, we have merely adapted the standard Taylor rule to our setting, introducing three monetary policy parameters $\rho^*$ (the natural interest rate)
and $\phi_\pi,\phi_\varepsilon$ (the Taylor rule parameters), and two targets: inflation $\pi^*$ and employment $\varepsilon^*$.
\item For the banking sector, we have essentially used trivial accounting rules to ensure a no-profit condition on loans and deposits. The only new parameter is $f$
and determines how much of the cost of bankruptcy must be paid by loans or by deposits. However, this parameter plays very little role in the qualitative 
behaviour of the economy, so we set it to $f = 1/2$ henceforth. 
\item For the households, we assume an Euler-like behaviour of the consumption budget as a function of the real interest rate on deposits, i.e. the consumption budget 
decreases when the real rate is high and increases when it is low, with a slope given by parameter $\alpha_c$.
\item For the firms, the hiring/firing and wage policy is affected by the real rate on loans. When it is high, indebted firms are more careful, firing more rapidly and 
hiring less. The unique parameter coupling the real interest rate to the firms policy is $\alpha_\Gamma$. 
\end{itemize}
These last two additional rules are to some extent arbitrary. However, they capture effects that certainly exist in the real world; furthermore changing the
detailed implementation we chose here while keeping the spirit of these rules lead to very similar conclusions. This is in line with our general claim that 
macro-behaviour is to a large degree insensitive to micro-rules. 
}

{
In summary, we want to emphasize that our extension of Mark-0 is very parsimonious: we have only added {\it two} behavioral parameters 
($\alpha_c,\alpha_\Gamma$). 
}

\subsubsection{Recovering Mark-0}

From the above discussion, we see that the core Mark-0 model of~\cite{Tipping} is recovered whenever the baseline interest 
rate is zero $\rho^*=0$, the CB is inactive ($\phi_\pi=\phi_\varepsilon=0$), and households and firms are insensitive to 
interest rates and inflation (i.e. setting $\alpha_c=\alpha_\Gamma=0$).

There is however a slight remaining difference with Mark-0 in the resolution of bankruptcies. The closest one can get is by setting $f=0$, 
i.e. absorbing default costs only through savings. In this case the only non-zero interest rate remaining in the dynamics of the model 
is the one on deposits, which is negative: $\rho^{\text{d}} = - {\DD(t)}/{\XX(t)} \leq 0$. 
This indeed roughly corresponds to the default resolution described in~\cite{Tipping} where default costs are paid by households 
and firms savings.\footnote{To be more precise, in the default resolution described in~\cite{Tipping} we introduce a bailout probability, called $f$ there, which 
sets the relative impact of default costs on households and firms savings. 
In this sense, the present setting recovers the one in~\cite{Tipping} with $f \approx 1/2$.}
There are also minor differences in the time-line of the model (in particular bankruptcies are resolved before price, production and wages are updated). 
All these differences however have a negligible quantitative impact on the results below.

\section{The ``natural'' behavior of the economy (without monetary policy)}
\label{results}

In this section we {analyze} the features of the model that arise from the introduction of interest rates in the Mark-0 economy, 
disregarding for a while any active monetary policy (i.e. setting $\phi_\pi=\phi_\varepsilon \equiv 0$ in Eq. (\ref{taylor_rule}) above).
In other words, we study an economy where the baseline interest rate is equal to $\rho^*$, constant in time, and affects both the 
consumption propensity of households through the parameter $\alpha_c$ appearing in Eq. (\ref{cons_budget}), and the firms hiring/firing 
propensity through the parameter $\alpha_\Gamma$ appearing in Eq. (\ref{alpha_Gamma}), with $\Gamma_0=0$ henceforth. 

When interest rates do not feedback at all into firms' and agents' {behavior} (i.e. for $\alpha_\Gamma=\Gamma_0=0$ and $\alpha_c=0$) the phenomenology of the Mark-0 model
is basically unaffected; in particular the phase diagram Fig.~\ref{fig:PD_wages} is unchanged. 

\subsection{Coupling between interest rates and firms {behavior}}

When $\alpha_\Gamma > 0$, on the other hand, 
the overall {behavior} of the economy evolves as expected: as long as $\rho^*$ is less than the inflation rate, nothing
much happens, in particular because Eq. (\ref{alpha_Gamma}) gives $\Gamma = \Gamma_0 \equiv 0$ here. When $\rho^*$ exceeds the
inflation rate, one observes that the unemployment rate $u$ starts increasing with $\rho^*$, while the demand for
credit and the inflation rate itself nosedive as expected. 

\begin{figure}
\centering
\includegraphics[scale=0.35]{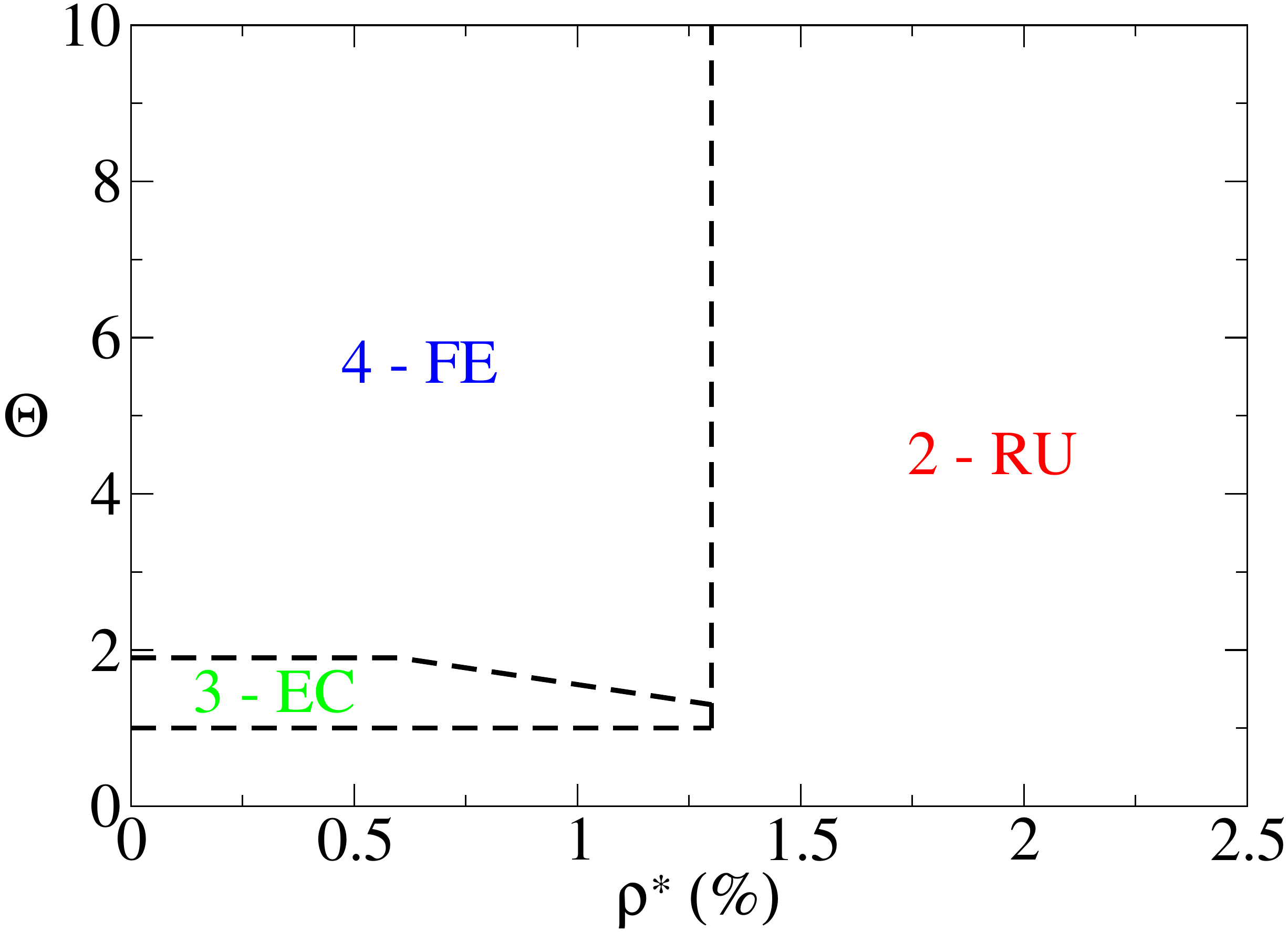}
\includegraphics[scale=0.35]{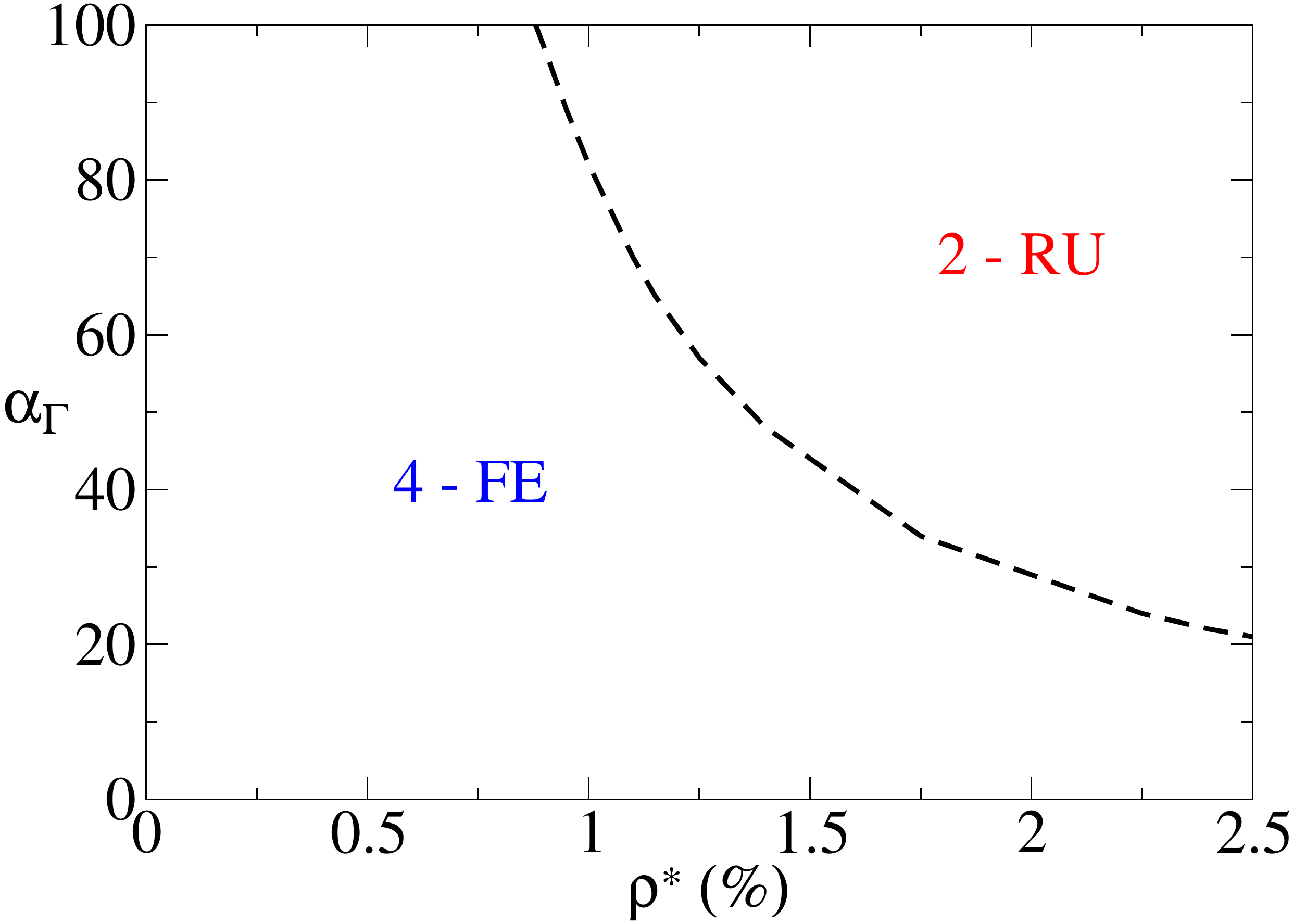}\\
\caption{
\emph{Left:} Phase diagram of the ``natural'' state of the economy in the $\rho^* - \Theta$ plane (i.e. with no monetary policy, $\phi_\varepsilon=\phi_\pi=0$) 
when both firms' and households' decisions are sensitive to the rate level: $\alpha_\Gamma= 50$ and $\alpha_c=4$.
For small enough rates $\rho^* < \rho^{**}$, one recovers the FE-EC-RU phases as $\Theta$ is decreased. When $\rho^*> \rho^{**}$, however, the RU phase prevails for all 
values of $\Theta$, which becomes irrelevant. 
\emph{Right:} Phase diagram in the $\rho^*-\alpha_\Gamma$ plane for $\Theta=3$, showing the dependence of the critical value $\rho^{**}$ on $\alpha_\Gamma$.
The other parameters of the model are set to: $R=2$ (with $\eta^0_-=0.1$), $c_0=0.5$, $\Gamma_0=0$, $\varphi=0.1$, 
$\gamma_p=\gamma_w=0.05$, $\beta=2$, $\delta=0.02$, $f=0.5$, $N_F=2000$ (see Appendix A for a definition of all these parameters).\\
}
\label{fig:th-rho-G}
\end{figure}

In Fig.~\ref{fig:th-rho-G} (left) we plot the phase diagram of the model in the $\rho^*-\Theta$
plane when $R = 2 > R_c$, and for a fixed value $\alpha_\Gamma= 50$.\footnote{This value of $\alpha_\Gamma$ has the following interpretation: when the 
debt of a firm equals its payroll, i.e. when $\Phi_i = 1$, a real interest rate of $8 \%$ annual leads to a freezing of all hires and a doubling of the 
firing rate, compared to a zero-debt situation.} For $\rho^*$ smaller than a certain value $\rho^{**} \approx 1.3$, one observes the familiar three phases FE-EC-RU as $\Theta$ is decreased, as
in Fig.~\ref{fig:PD_wages}. However, for a baseline rate larger than $\rho^{**}$, the FE and EC phase disappear entirely, and the Residual Employment phase
(with $u \sim 30 \%$) prevails for all values of $\Theta$. To wit, when interest rates are too high, firms hesitate to accumulate more debt (even if they are allowed to 
when $\Theta$ is high). Rather, they prefer reducing their work force when needed, keeping the unemployement at a relatively high level. 

Fig.~\ref{fig:th-rho-G} (right) allows one to understand the role of $\alpha_\Gamma$: there 
we show the phase diagram in the $\rho^*-\alpha_\Gamma$ plane for fixed values of $R=2$ and $\Theta = 3$, such that the economy is in the Full Employment phase for $\alpha_\Gamma=0$. 
A sudden phase transition between the FE and RU phases occurs for a critical value $\rho^{**}(\alpha_\Gamma)$; the
larger the sensitivity to interest rates -- i.e. the larger $\alpha_\Gamma$ -- the smaller the critical value of the baseline interest rate beyond 
which the economy is {destabilized}. It is interesting (and quite counterintuitive) that the {\it aggregate {behavior} of the economy is not a smooth function of the
interest rate}. When firms are risk averse and fear going into debt, large enough interest rates lead to more unemployment that spirals into a {destabilizing} feedback loop. 
This is one of the ``dark corners'' that ABMs can help uncovering.

\subsection{Coupling between interest rates and household {behavior}}

Perhaps unexpectedly, the coupling between interest rate and consumption (captured by parameter $\alpha_c$) appears to have much smaller influence on the 
``natural'' state of the economy, at least when $\alpha_c$ is chosen 
within a reasonable range. Its main influence is to increase the output fluctuations around the steady state, by amplifying price trends 
through the resulting reduction/increase in consumption. Interestingly, we find that for $\Theta \gg 1$ and independently of $\rho^*$, 
$\alpha_c$ has a stabilising effect on the economy: $R_c$ shifts to lower values as $\alpha_c$ increases (in the absence of monetary policy). 
Clearly, micro-data is needed to estimate the value of $\alpha_c$ for realistic applications; but since this parameter plays a small role we will choose rather arbitrarily 
$\alpha_c=4$, unless explicitly stated. 
This corresponds to a moderate sensitivity to inflation/interest rates: a rise of the interest rate of $1 \%$/year  
corresponds to a decrease of $4 \%$ of the consumption propensity. 

\section{Monetary policy experiments}

We now consider a simple policy framework where the Central Bank adjusts the base interest rate $\rho_0(t)$ in order to
achieve its inflation and employment targets. Given the simplicity of our model we are mainly interested here in gaining a qualitative
understanding of the possible consequence of a Taylor-rule based monetary policy in a stylized Agent-Based framework. We defer parameter calibration, 
detailed comparison with simple DSGE models and more quantitative insights to future studies. 

Our major finding is that 
provided its policy is not too aggressive and the economy not too close to a phase transition, the CB is successful in steering the
economy towards its targets. However, the mere presence of different equilibrium states of the economy separated by phase boundaries 
(i.e. ``dark corners'') may deeply alter the impact of monetary policy. Indeed, we will exhibit cases where
the monetary policy {\it by itself} triggers large instabilities and is counter-productive.  

As in the previous section, we will refer to the state obtained without any  
response of the CB, i.e. when $\phi_\pi=\phi_\varepsilon=0$, as the ``natural'' state of the economy (for a given set of parameters).
The corresponding ``natural'' value of a variable $x$ will be denoted by $x_{\text{nat}}$.
In order to simplify the analysis and since most of the parameters of our model play little role in the qualitative
{behavior} of the economy we set once and for all some of them to the values given in the caption of Fig.~\ref{fig:th-rho-G} and 
choose $\rho^*=2 \%$. We only focus on the four parameters defining the CB policy (i.e. $\phi_\pi,\ \phi_\varepsilon,\ {\varepsilon}^*$ and ${\pi}^*$), 
the parameters of the transmission channels (i.e.  $\alpha_c$ and $\alpha_\Gamma$), and the two parameters 
locating the system in the phase diagram of Fig.~\ref{fig:PD_wages} (i.e.  the hiring/firing ratio $R$ and the bankruptcy threshold $\Theta$). 

\begin{figure}
\centering
\includegraphics[scale=0.45]{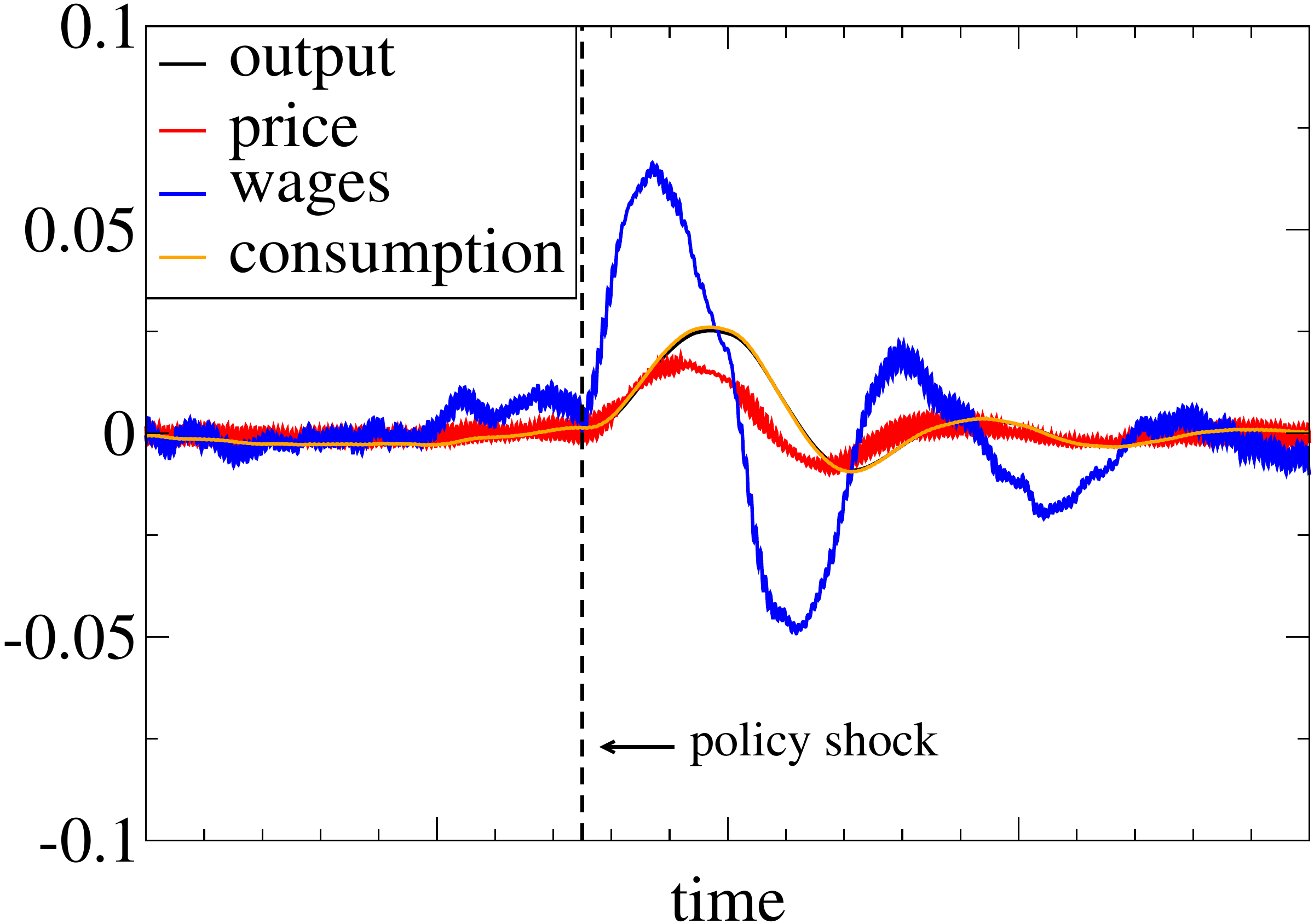}
\caption{Aggregate response of the economy to a policy shock represented by a drop in the interest rate $\rho_0$ set by the central bank (from $2\%$ to $1.8\%$).
The vertical dashed line gives the timing of the policy shock. All aggregate variables are represented as a relative variation from their average values prior to
the shock. For a better visualization, we amplified by a factor $100$ the wage and price responses.  
Note that the damped oscillation relaxation towards equilibrium. 
Parameters are $\Theta=3$, $\rho^*=2\%$, $\alpha_\Gamma = 50$, $\alpha_c=4$, and other parameters as in Fig. \ref{fig:th-rho-G}.
}
\label{fig:shock}
\end{figure}

\subsection{A monetary shock}

{
In order to check that our framework produces reasonable results, it is interesting to start with the simplest policy experiment, 
i.e. an expansionary monetary policy shock where the baseline interest rate is 
instantaneously decreased from  $2 \%$ to $1.8 \%$, all other parameters of the model being fixed, and with no further intervention of the Central
Bank (i.e. $\phi_\pi = \phi_\varepsilon=0$). The resulting dynamics of output, wages and prices is shown in 
Fig.~\ref{fig:shock} and can be compared with e.g. Fig 1 of Ref.~\cite{Christiano}. 
It is gratifying to see that our simple modelling strategy, in particular the way the monetary policy is channelled to the economy, 
does lead to a quite realistic shape of impulse response functions, at least qualitatively. 
In fact, the empirical hump-shaped response of the output reported in~\cite{Christiano} is quite well reproduced by their DSGE model with nominal rigidities. 
Although the underlying microfundations are very different, our 
ABM indeed includes frictions: all adjustments (of production, wages and prices) in our model are only made progressively. So 
our framework is indeed able to capture effects predicted by enhanced DSGE models. 
No attempt was made here to adjust parameters to quantitatively match empirical data, in particular the time scales:
this is beyond the scope of the present study which is primarily intended to be a description of the qualitative behavior of our toy-economy.
}

\subsection{Mild vs aggressive monetary policy}

\begin{figure}
\centering
\includegraphics[scale=0.3]{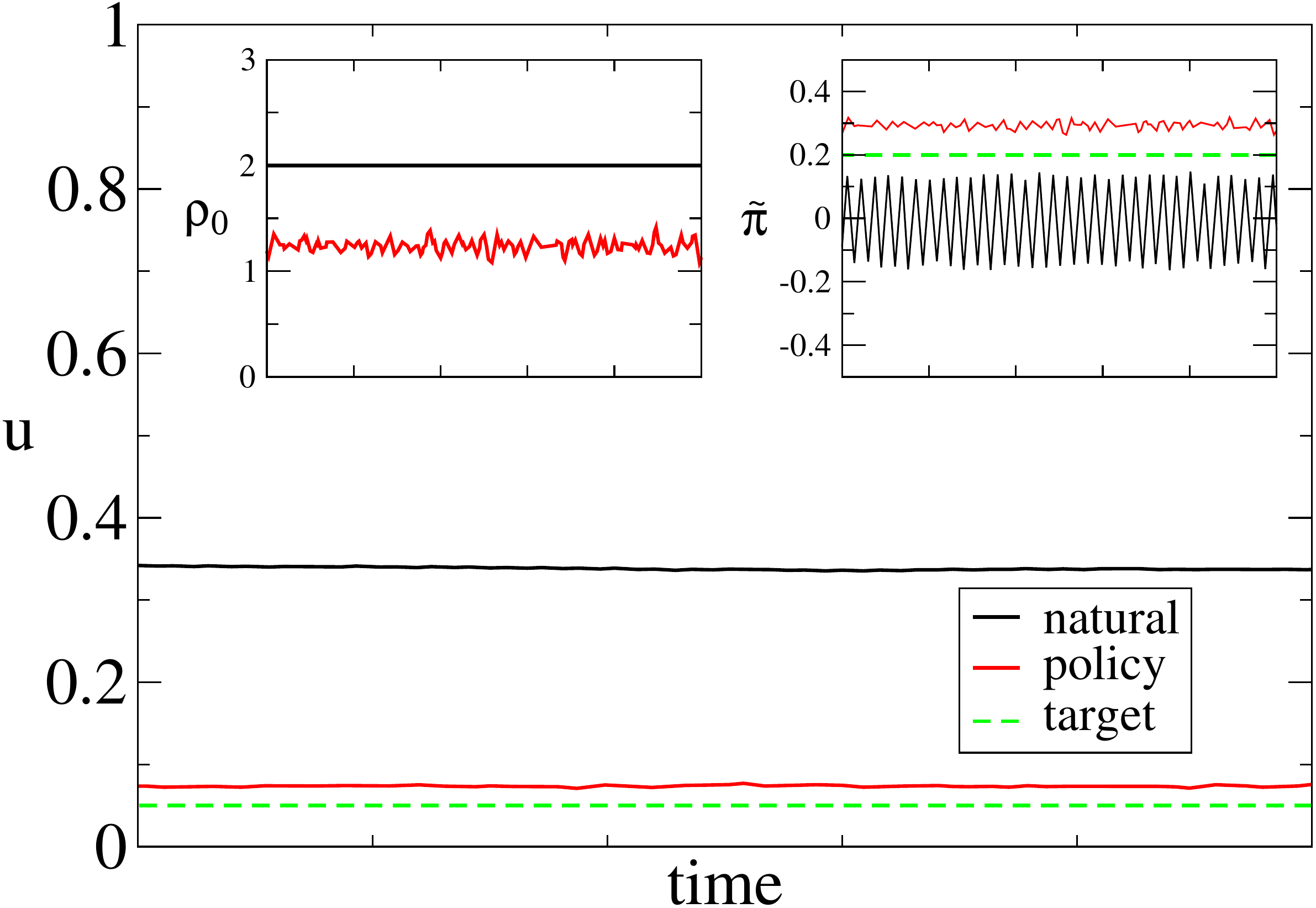}
\includegraphics[scale=0.3]{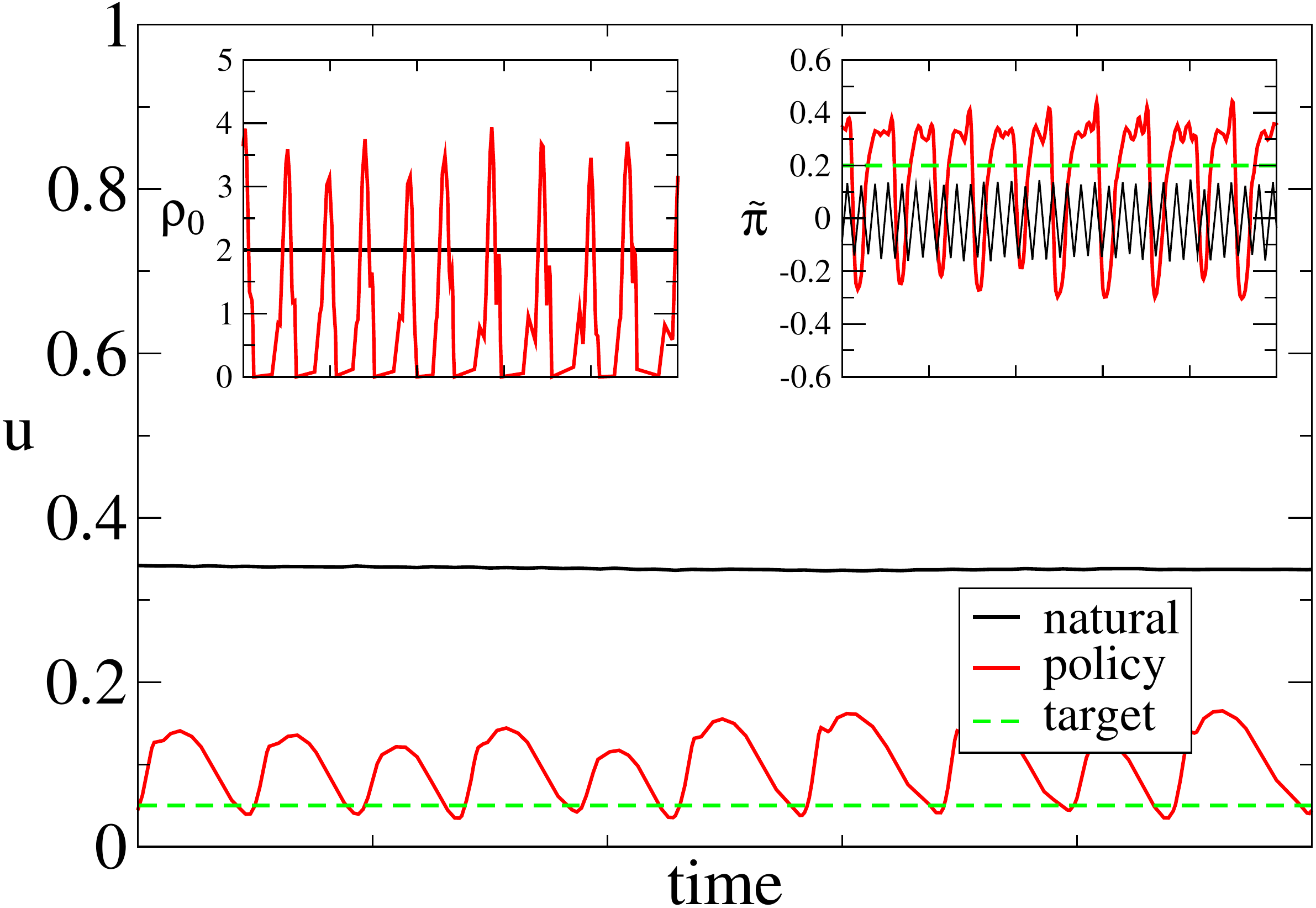}\\
\caption{
In these plots we select the same natural state of the economy with $\Theta=2$, $R=2$, $\rho^*=2\%$ and $\alpha_\Gamma = 50$, such that
$u_{\text{nat}} \approx 0.33$ and $\pi_{\text{nat}} \approx 0$ (but fluctuating). 
The CB sets a much lower unemployment target of $u^* = 0.05$ and an inflation 
target of ${\pi}^* = 0.2 \%$. In the left plot the CB policy is moderate ($\phi_\pi=\phi_\varepsilon=0.5$) and basically achieves its goals, while in the  
right plot the policy is aggressive ($\phi_\pi=\phi_\varepsilon=1$) and {destabilizes} the whole economy. 
In the insets, $\r_0$ and $\pi$ are given in $\%$s.
The period of the business cycles in the right plot corresponds to $\approx 4$ years 
if the elementary time unit of the model is one month.
Larger spikes of unemployment are also observed with much lower frequency (not shown).
}
\label{fig:policy-runs}
\end{figure}

In Fig.~\ref{fig:policy-runs} for example, we show the result of the policy of the Central Bank that attempts to bring down the natural unemployment 
rate of $u_{\text{nat}} \approx 0.33$ (a rather large value corresponding to $\Theta=2$, $R=2$ and $\rho^* = 2\%$) to a low target of $1 - {\varepsilon}^* = u^* = 0.05$. 
The target inflation is ${\pi}^* = 0.2 \%$ per time step (corresponding to $2.4 \%$ annual if one interprets the time step to be a month), compared to a natural 
inflation that fluctuates around zero: $\pi_{\text{nat}} \approx 0$. The left graph corresponds to
a mild monetary policy, with Taylor-rule parameters set to $\phi_\pi=\phi_\varepsilon=0.5$. The policy is seen to be rather successful: the inflation is 
on target, while unemployment goes down to $u \approx 0.07$, not far from the target of $0.05$. 
But now look at the graph on the right, where the only difference is the aggressiveness
of the CB that attempts to reach target too quickly, merely doubling the value of $\phi_\pi=\phi_\varepsilon \to 1$. In this case, the 
monetary policy has induced strong instabilities, with ``business cycles'' of large amplitude and inflation all over the place. 

\begin{figure}
\centering
\includegraphics[scale=0.6]{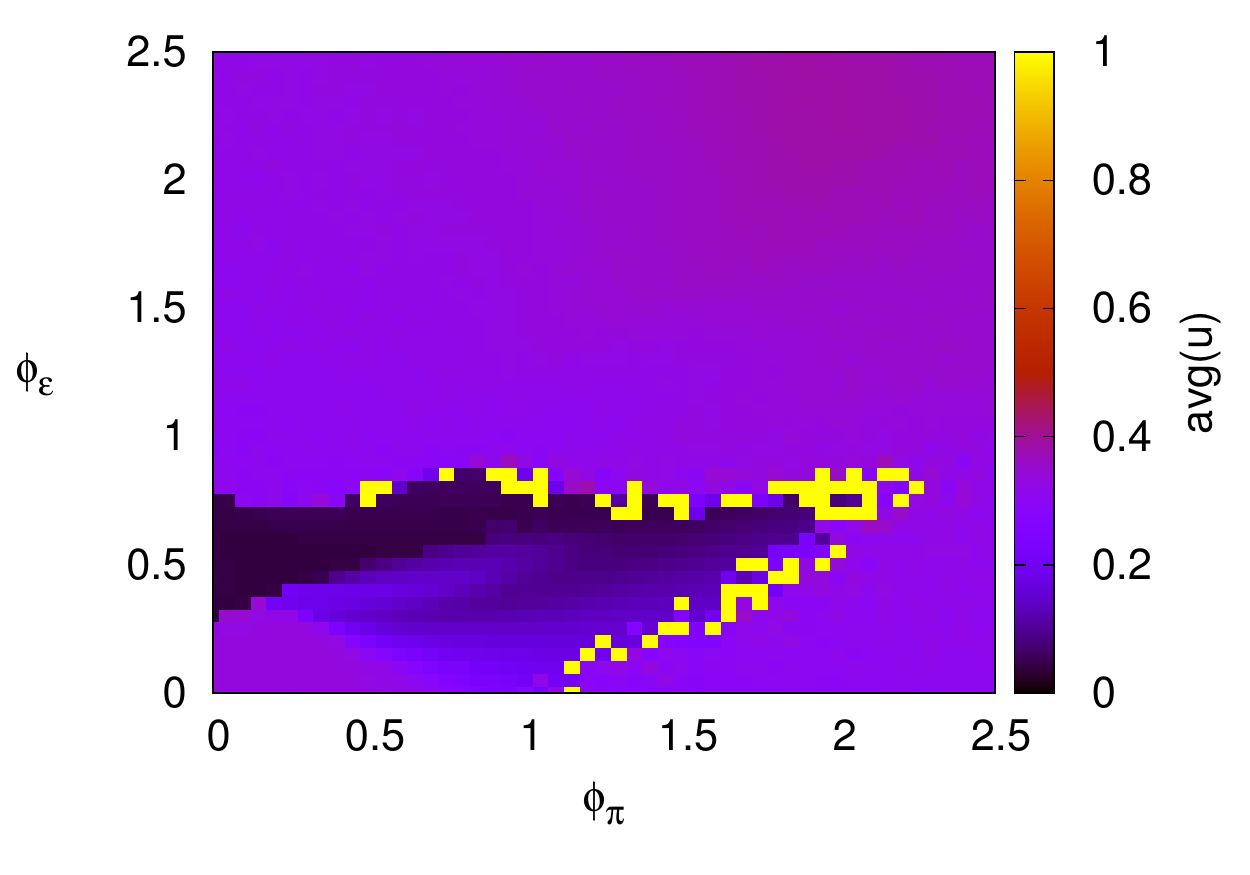}
\includegraphics[scale=0.6]{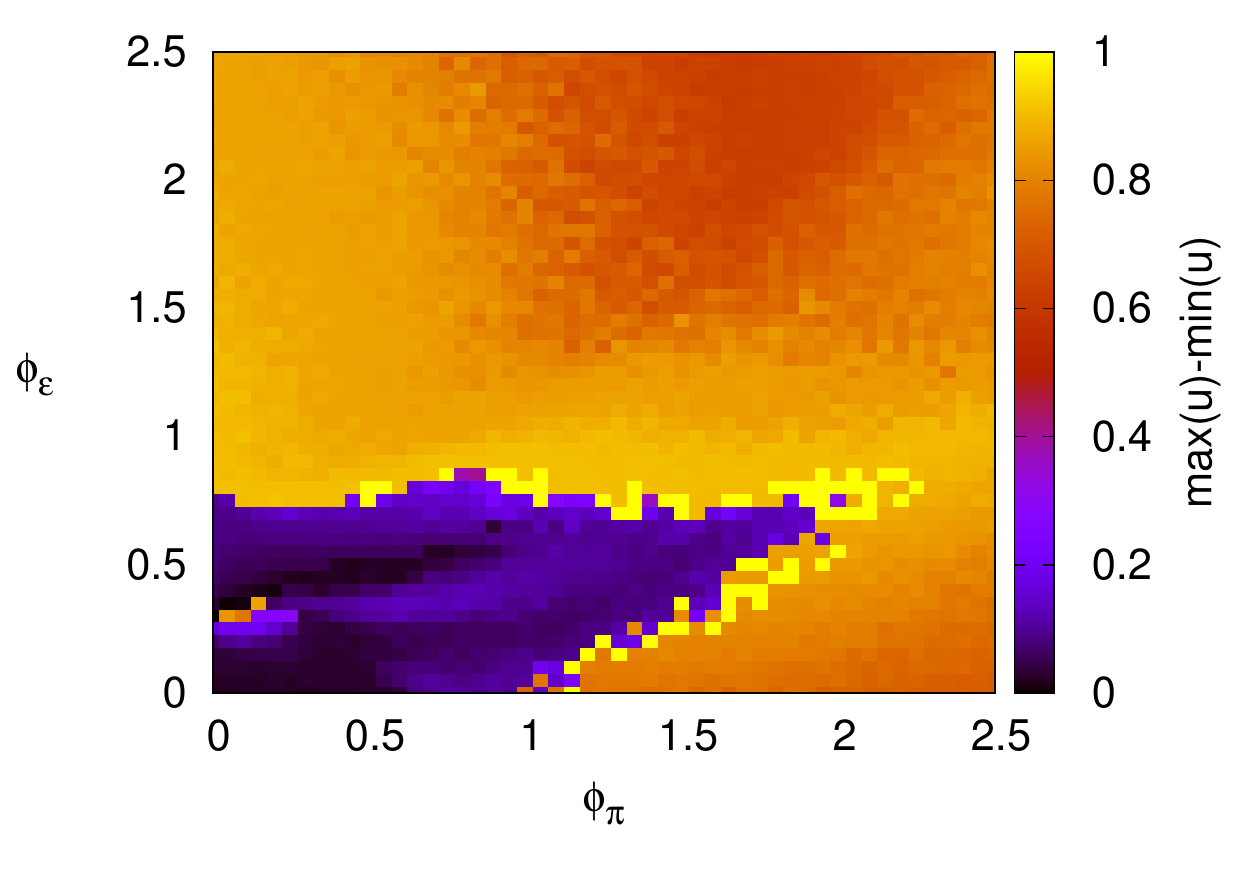}
\caption{
Policy performance in the $(\phi_\pi,\phi_\varepsilon)$ plane (the origin corresponds to the natural state of the
economy), for $\Theta=2$, $R=2$, $\rho^* = 2\%$, $\alpha_\Gamma=50$ and $\alpha_c=4$. The target inflation and unemployment are, respectively, 
$\pi^*= 0.2 \%$ and $u^*=0.05$. \emph{Left:} Color map of the average unemployment; \emph{Right:} Color map 
of the amplitude of the business cycle, measured as $\max_t(u)-\min_t(u)$. Yellow/orange regions correspond to unstable economies 
with crises of large amplitude. 
As one can see the policy is effective as long as it is not too aggressive, with a sharp transition to a regime
where it become{s} detrimental. The closer the natural economy to a phase boundary, the more {destabilizing} the policy -- see Fig.~\ref{fig:th-R-agg}.
}
\label{fig:ap-ae}
\end{figure}

We show in Fig.~\ref{fig:ap-ae}
the result of an extensive exploration of the role of $\phi_\pi$ and $\phi_\varepsilon$ when $\Theta=2$, $R=2$ and $\rho^* = 2\%$. One sees that
there is a wedge-like region around $\phi_\pi=\phi_\varepsilon=0$ where the policy does not induce instabilities 
({signaled} by a yellow/orange hue in Fig.~\ref{fig:ap-ae}-right). However, the region of parameters where the unemployment rate is significantly reduced is 
only a subset of this wedge, corresponding to the black region in Fig.~\ref{fig:ap-ae}-left, where $\phi_\varepsilon \sim 0.5$, $\phi_\pi \leq 0.5$. In other words, 
the Central Bank must navigate in a narrow window: too little is not enough, too aggressive is counterproductive and leads to instabilities and wildly 
oscillating economies. 

Fig.~\ref{fig:ap-ae} also reveal that in our toy-world, a Central Bank with a dual mandate (inflation and output) enjoys a wider region of stability compared to a Central
bank with inflation as the only mandate. Steering inflation and output simultaneously increases the probability of a successful monetary policy.

\subsection{The role of phase boundaries}

\begin{figure}
\centering
\includegraphics[scale=0.6]{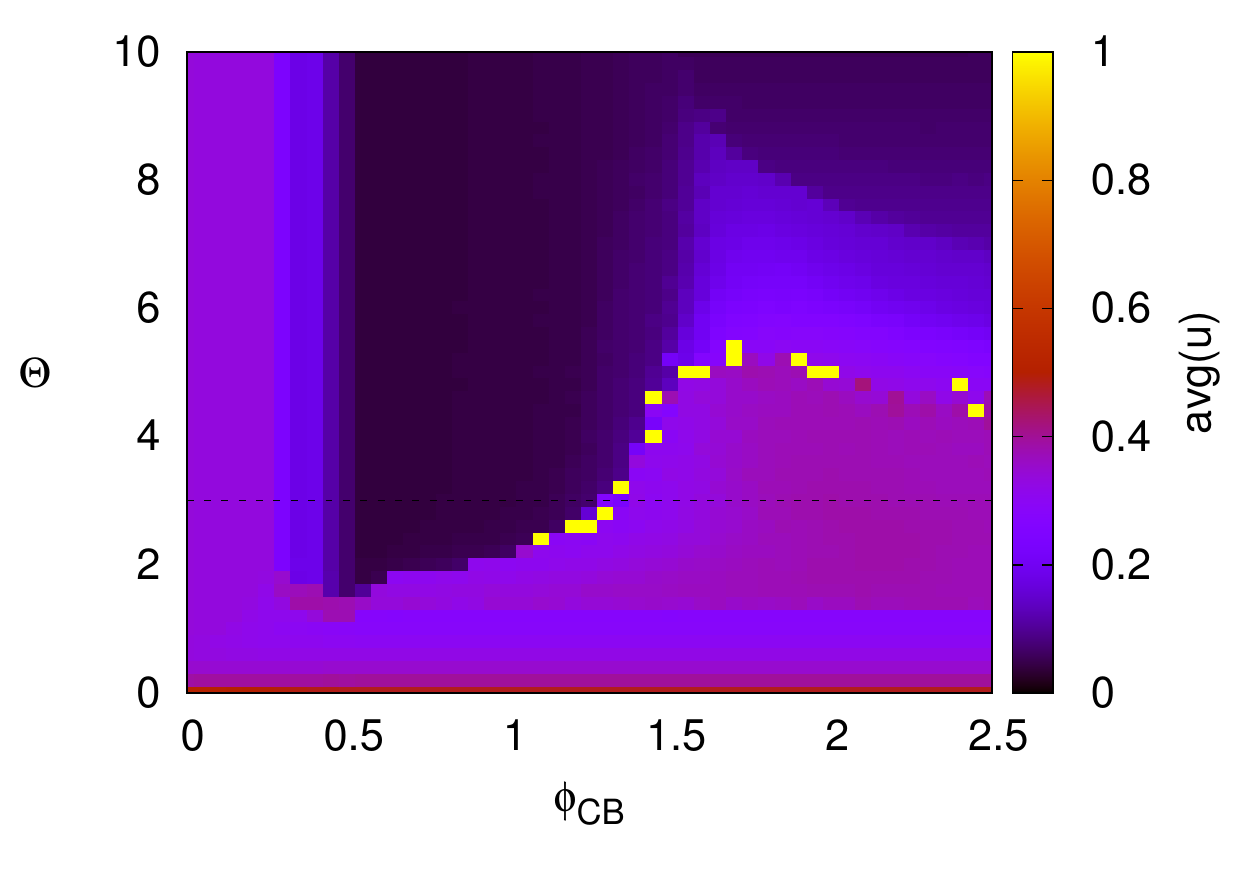}
\includegraphics[scale=0.6]{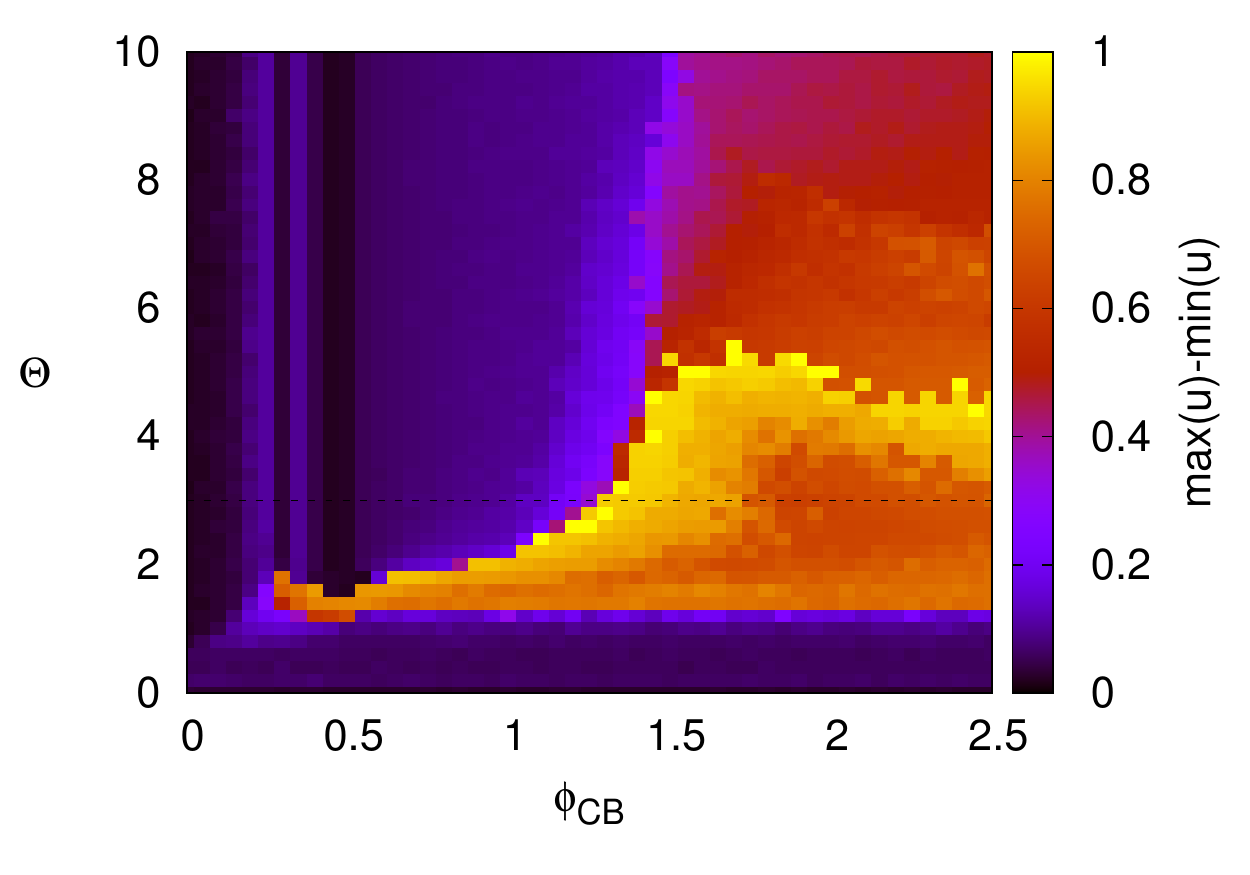}\\
\includegraphics[scale=0.6]{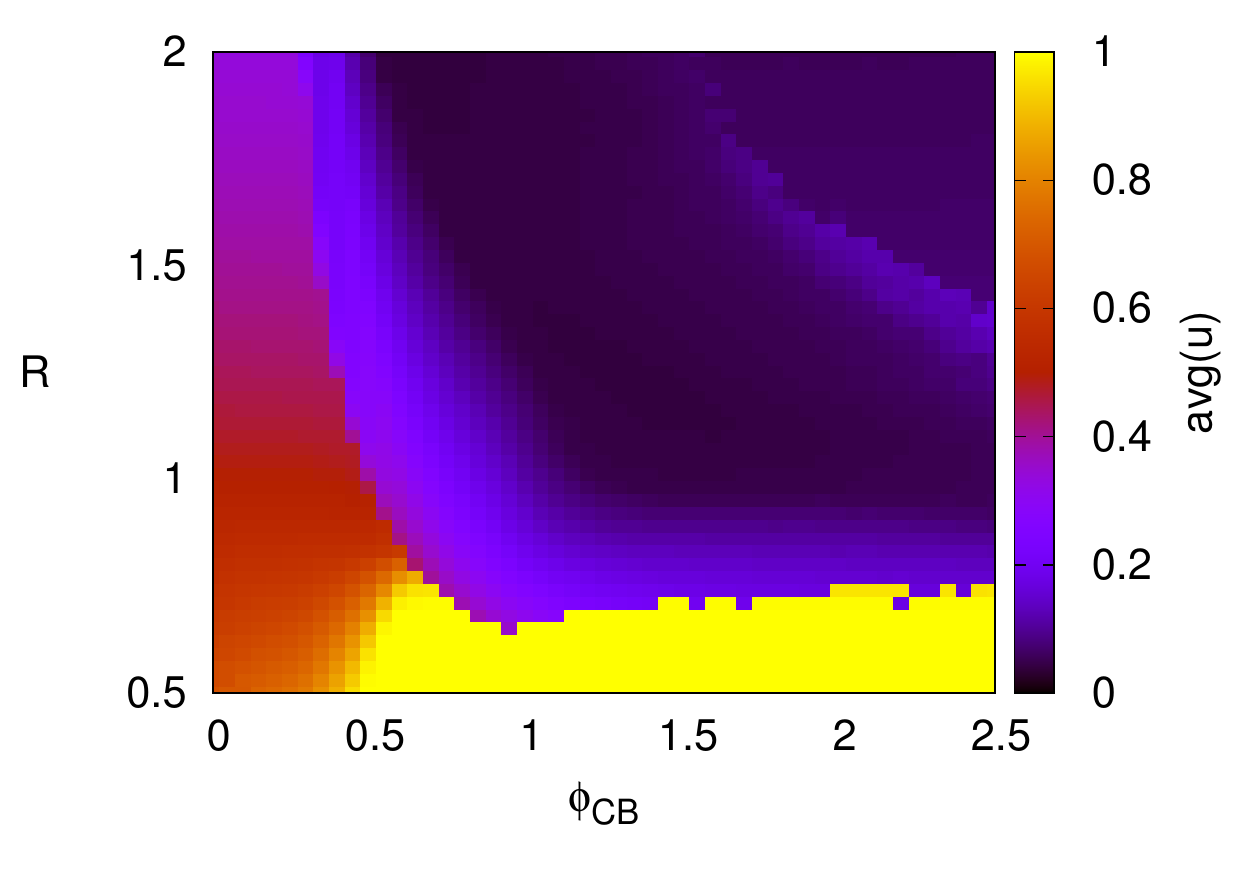}
\includegraphics[scale=0.6]{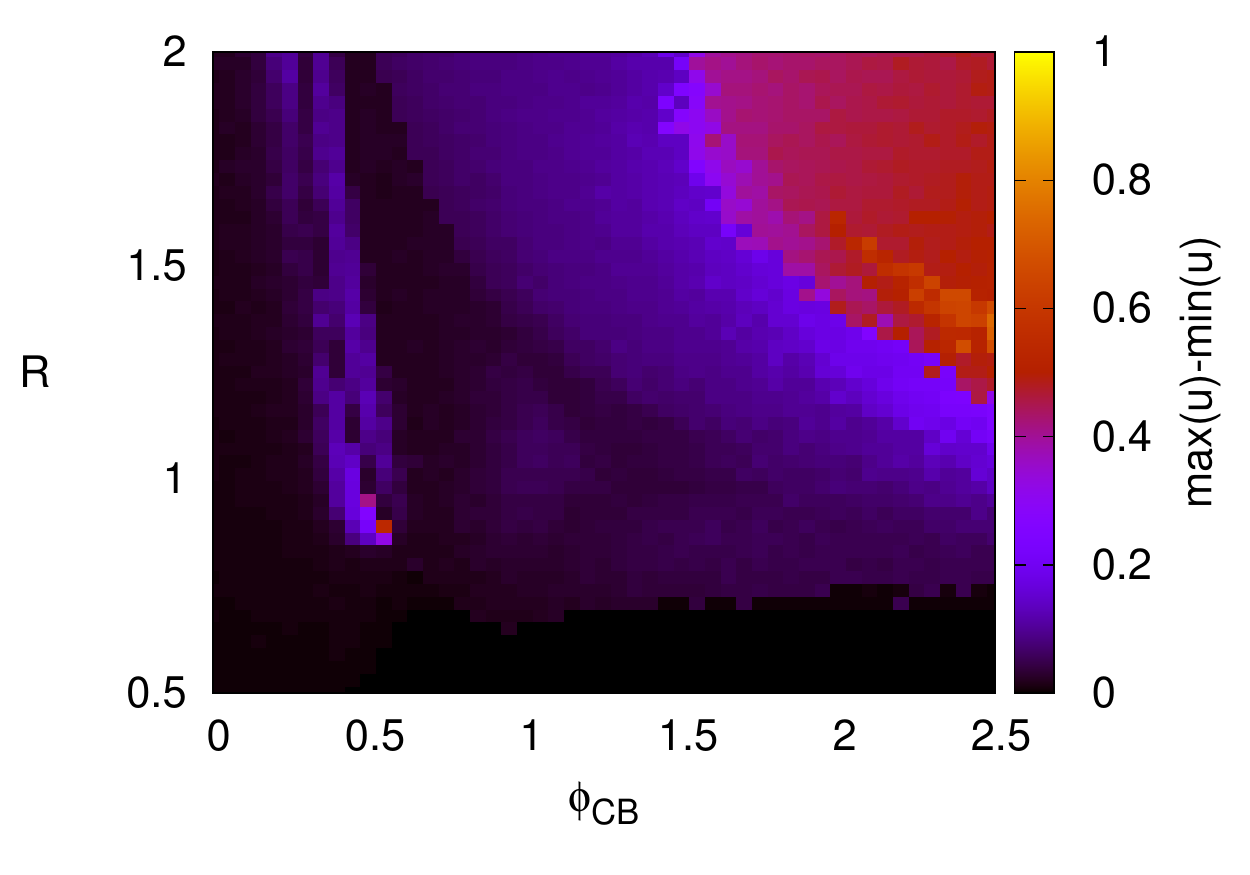}
\caption{A more systematic diagram of the effect observed in Fig.~\ref{fig:ap-ae}. Here we move on the diagonal
$\phi_\pi=\phi_\varepsilon$ of Fig.~\ref{fig:ap-ae} on the
$\phi_{\text{CB}}$-axis and plot the average unemployment $u$ and unemployment variations  $\max_t(u)-\min_t(u)$. The other axis is either $\Theta$ (top row, for $R=2$) or 
$R$ (bottom row, for $\Theta=10$), still with $\alpha_\Gamma=50$, $\alpha_c=4.$, $\pi^*= 0.2 \%$ and $u^*=0.05$.  Note that since $\alpha_\Gamma > 0$, 
the transition at $R_c \approx 1, \phi_{\text{CB}}=0$ is 
second order (see Appendix B). As above, yellow/orange regions in the plots on the right, correspond to unstable economies with crises of large amplitude.}
\label{fig:th-R-agg}
\end{figure}

The fragility of the economy is clearly due to the proximity of the phase boundary that appears in Fig.~\ref{fig:th-rho-G}. 
As one moves away from the boundary, 
for example by increasing $\Theta$, one finds that the region where the CB policy is harmful shrinks. We illustrate this
by moving along the line $\phi_\pi=\phi_\varepsilon \equiv \phi_{\text{CB}}$ in parameter space. We display 
in  Fig.~\ref{fig:th-R-agg} the phase diagram of the model in the $(\phi_{\text{CB}},\Theta)$ plane and in the $(\phi_{\text{CB}},R)$ plane. 
One clearly sees from the left graph on the top row that 
the deep blue region (corresponding to low unemployment) expands for larger $\Theta$, and that the yellow/orange region of the graph on the
right (corresponding to strong oscillations) recedes. 

The bottom graphs illustrate the role of $R$. One mostly observes that: 
\begin{itemize}
\item (a) deep in the FU phase ($R < 0.75$), 
the monetary policy is helpless in restoring employment; 
\item (b) for intermediate values of $R$, large enough values of $\phi_{\text{CB}}$ do lead to 
small unemployment rates; 
\item (c) when $R$ and $\phi_{\text{CB}}$ are simultaneously large, instabilities appear 
(see  Fig.~\ref{fig:th-R-agg} right bottom graph, north-east corner). 
\end{itemize}
Point (a) above is interesting and can be understood as follows: when $R$ is small, 
firms are so quick to adjust production downwards that they never need credit, and their financial fragility is low or even negative. 
The interest-rate impact parameter $\alpha_\Gamma$ then becomes completely ineffective in this case. 
This could be relevant to understand the aftermath of the 2008 crisis: if one interprets strong downsizing (i.e. small $R$) as a result of a drop of confidence induced
by the Lehman crisis, the above discussion suggests that a low interest rate policy might not be as effective as one may have expected. 

As a complement to the above analysis we also investigated the performance of the policy as a function of the distance 
between target and natural values. We find, not surprisingly, that when targets are not too far from the natural state of the economy the CB
manages to achieve its targets without triggering any instability. When however targets are far from the natural state even a mild policy may become
detrimental and trigger instabilities.

\subsection{The role of households and firms sensitivity to rates}

\begin{figure}
\centering
\includegraphics[scale=0.6]{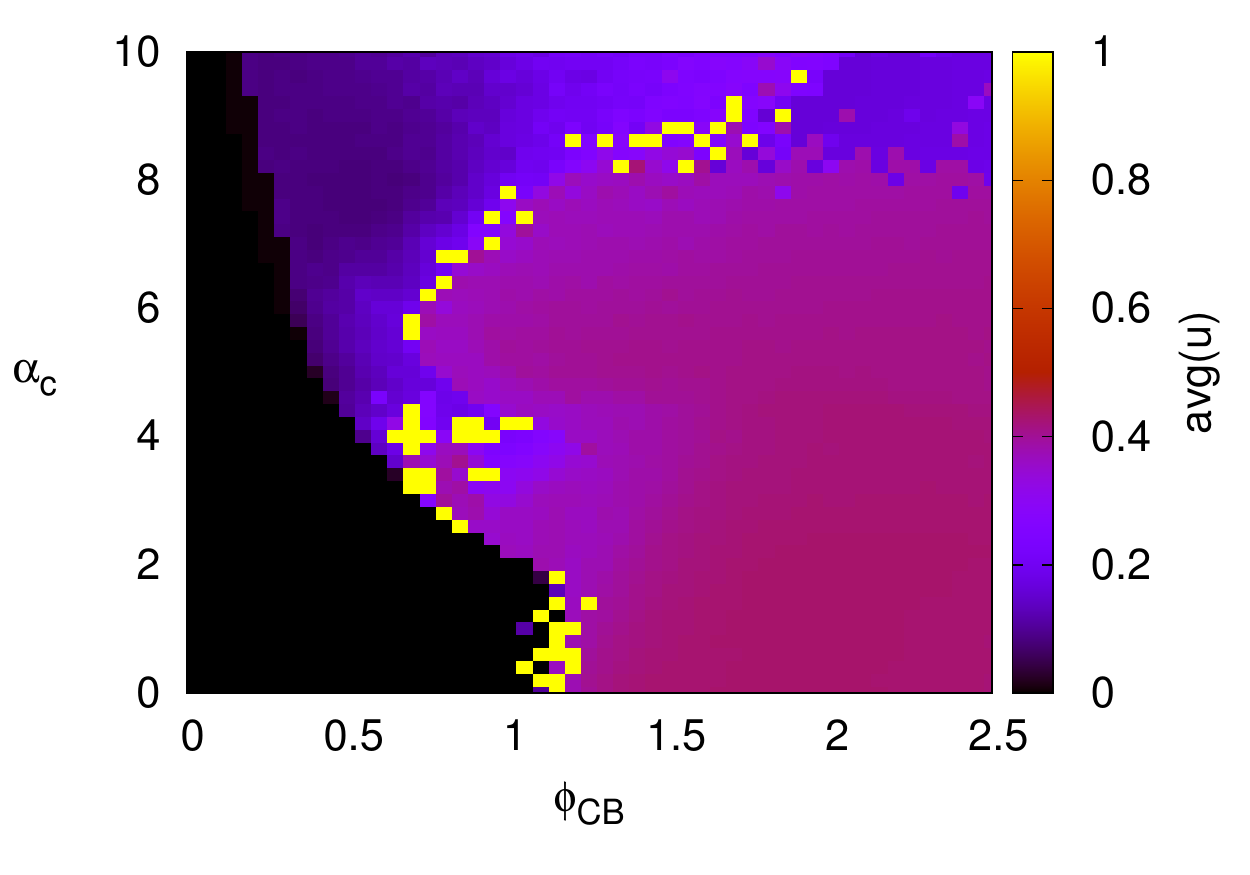}
\includegraphics[scale=0.6]{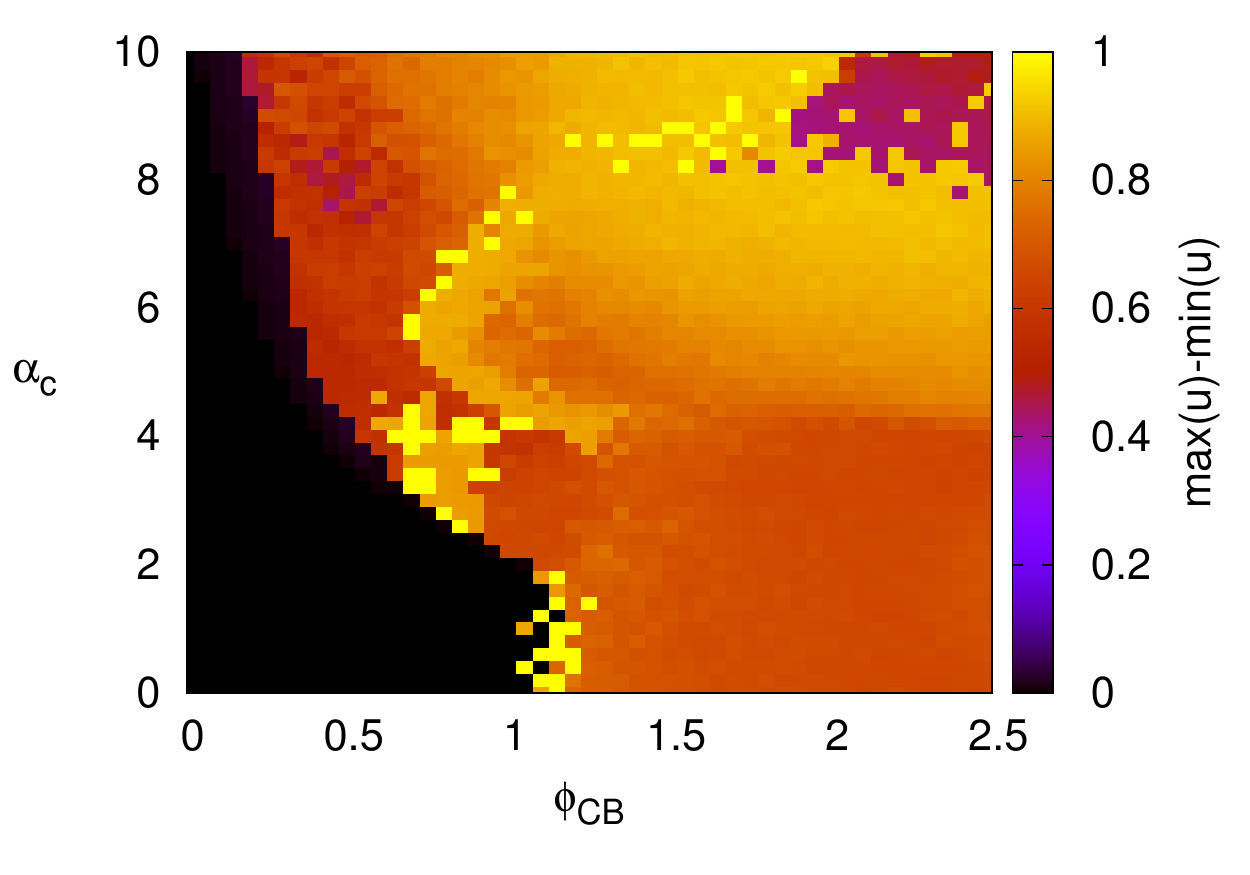}\\
\includegraphics[scale=0.6]{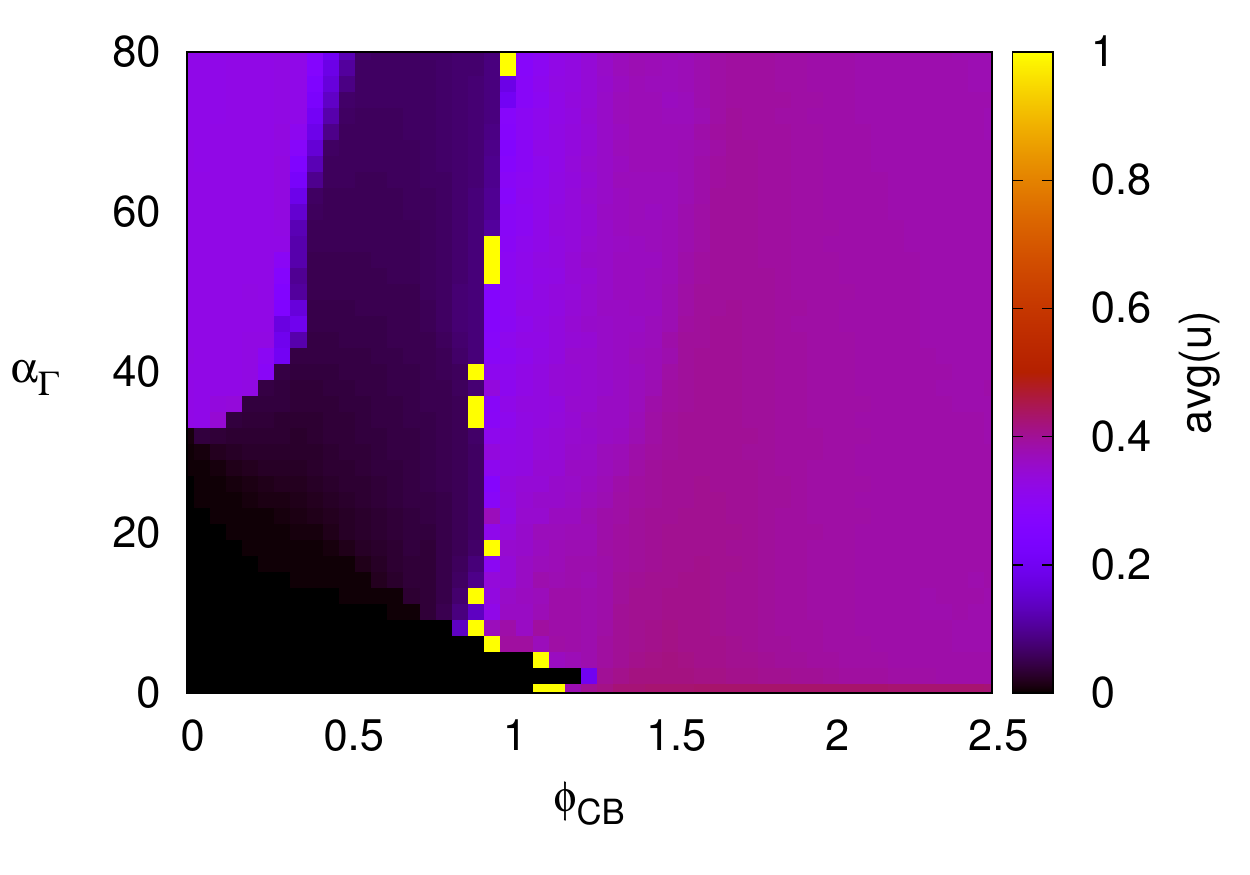}
\includegraphics[scale=0.6]{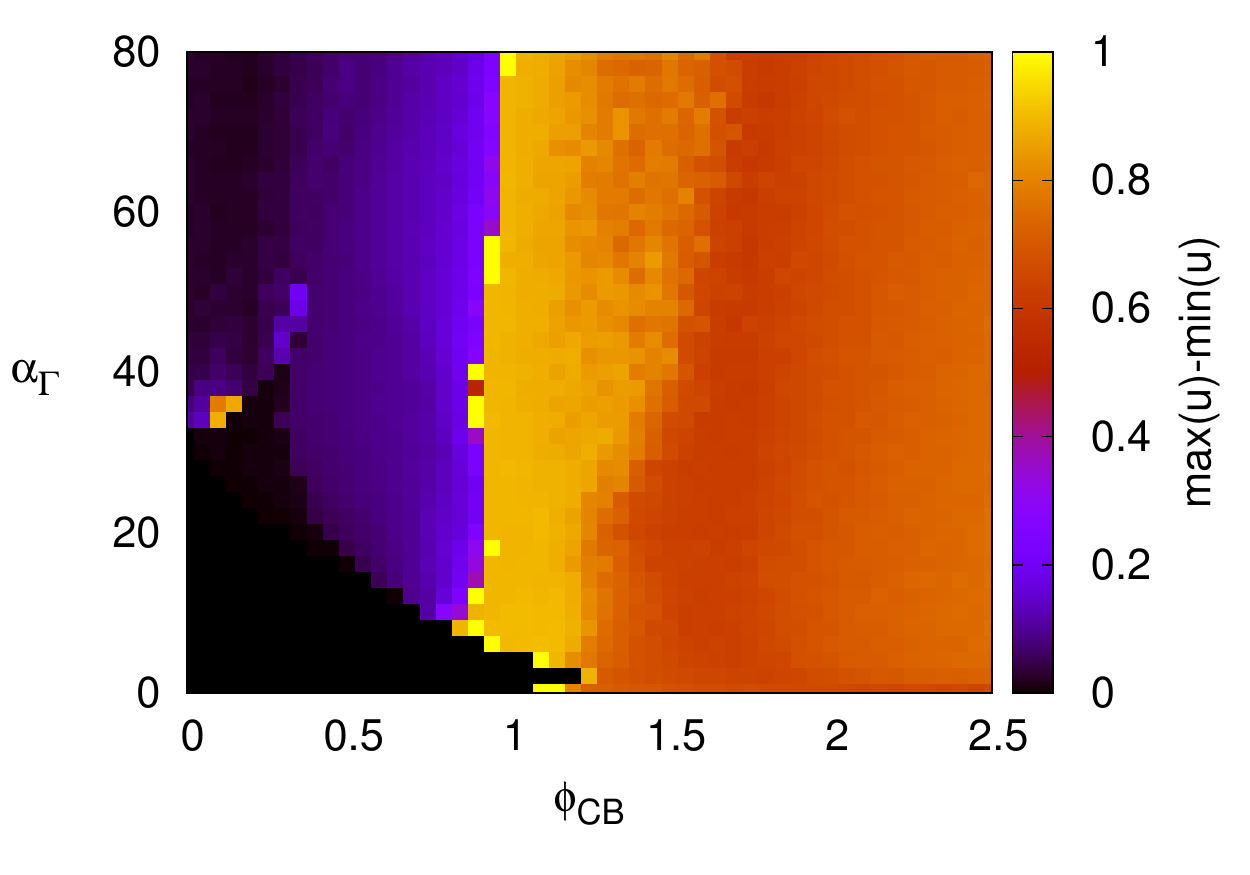}
\caption{Policy performance, as in the previous two plots, now separately as a function of  the households sensitivity
to rates $\alpha_c$ and the firms sensitivity to rates $\alpha_\Gamma$. The $x$-axis is again $\phi_\pi=\phi_\varepsilon \equiv \phi_{\text{CB}}$,
whereas $\Theta=3$, $R=2$ and $\rho^* = 2\%$, with again $\pi^*= 0.2 \%$ and $u^*=0.05$.
(\emph{Top row}) Here $\alpha_\Gamma=0$ and $\alpha_c > 0$. The FE (black) region prevailing for $\alpha_c=\phi_{\text{CB}}=0$ is destabilized beyond a certain
$\phi_{\text{CB}}^*$ that decreases as $\alpha_c$ increases. Note the constellation of yellow points that correspond to a completely broken economy. 
(\emph{Bottom row}) We now set $\alpha_c=0$ and vary $\alpha_\Gamma > 0$. When $\phi_{\text{CB}}=0$, one recovers the FE-RU transition  
for $\alpha_\Gamma \approx 30$ observed in Fig.~\ref{fig:th-rho-G}. 
One sees that for larger $\alpha_\Gamma$'s, the CB policy is at first successful in reinstalling the FE phase, before {destabilizing} again the economy beyond
the vertical line $\phi_{\text{CB}}^* \approx 0.9$.
}
\label{fig:ac-g0-gstar}
\end{figure}

We have also investigated the role of the two policy transmission channels ($\alpha_\Gamma$, firms and $\alpha_c$, households) separately, 
and found that both channels in isolation may trigger instabilities. In Fig.~\ref{fig:ac-g0-gstar} we plot the policy performance in the 
$(\phi_{\text{CB}},\alpha_c)$ plane and the $(\phi_{\text{CB}},\alpha_\Gamma)$ plane, with all other parameters fixed, in particular 
$R=2$ and $\Theta=3$. As expected, the top row shows that the larger the value of $\alpha_c$, the more careful the monetary policy has to be in order to 
avoid instabilities. Note, interestingly, that there is a thin region in the $(\phi_{\text{CB}},\alpha_c)$ plane (spot the yellow dots) where unemployment 
goes to $u \approx 1$, i.e. the economy is completely destabilized by the monetary policy! 

The bottom row of Fig.~\ref{fig:ac-g0-gstar} shows the interplay between policy aggressiveness $\phi_{\text{CB}}$ 
and firms sensitivity to the real interest rate $\alpha_\Gamma$. When $\phi_{\text{CB}}=0$, one recovers the FE-RU transition  
for $\alpha_\Gamma \approx 30$ already observed in Fig.~\ref{fig:th-rho-G}. For larger $\alpha_\Gamma$'s, 
the CB policy is at first successful in reinstalling the FE phase, before {destabilizing} again the economy beyond
$\phi_{\text{CB}}^* \approx 0.9$. When $\alpha_\Gamma$ becomes small, a {portion} of the FE region expands up to higher values of $\phi_{\text{CB}} \approx 1.25$.

A detailed comparison between Fig.~\ref{fig:th-R-agg} (top row) and Fig.~\ref{fig:ac-g0-gstar} (bottom row) reveals that the simultaneous 
presence of the two transmission channels has an overall {\it stabilizing} effect. Indeed, when only one channel is present 
(i.e. either $\alpha_c$ or $\alpha_\Gamma$ is zero) the value of $\phi_{\text{CB}}$ beyond which the system is unstable is {\it smaller} than 
when both channels are active. For example, for $\Theta=3$, $\alpha_c=4.$ and $\alpha_\Gamma=50.$ one sees from Fig.~\ref{fig:th-R-agg} (top row; left, horizontal line) that
the economy is destabilized for $\phi_{\text{CB}}^* \approx 1.3$, to be compared to $\phi_{\text{CB}}^* \approx 0.9$ when $\alpha_c=0$ and to $\phi_{\text{CB}}^* \approx 0.5$ 
when $\alpha_\Gamma=0$. Although we cannot provide a complete interpretation, this interesting observation might be due to the fact that the $\alpha$ parameters 
can be interpreted as ``awareness'' parameters that allow firms and agents to better anticipate crises and thus dampen the destabilising influence of the Central Bank policy.

\subsection{Comparison with DSGE models}

It is quite interesting to compare the diagram of Fig.~\ref{fig:ap-ae} with its DSGE counterpart (see e.g~\cite{DSGE}, section 4.3). Within our Agent-Based framework, small
values of $\phi_\pi,\phi_\varepsilon$ are at worst ineffective wheras large values of $\phi_\pi,\phi_\varepsilon$ lead to instabilities. This is at the 
opposite of what is observed in the fully rational world of DSGE models, where {\it small} values of $\phi_\pi,\phi_\varepsilon$
lead to instabilities, while large values of $\phi_\pi,\phi_\varepsilon$ allow the Central Bank to funnel the economy on a stability path~\cite{DSGE}. 
The intuition for this is made crystal clear  by Gal\`i in~\cite{DSGE}: {\it The monetary authority should respond to deviations of
inflation and the output gap from their target levels by adjusting the nominal rate
with ``sufficient strength''; [...] it is the presence of a ``threat'' of a strong response by the monetary authority to an eventual deviation of
the output gap and inflation from target that suffices to rule out any such deviation
in equilibrium.} In other words, fully rational, forward looking agents know that inflation will be tamed by the response of central authorities, {an}
element that is indeed completely absent in the myopic forecast world of most ABMs (including Mark-0), where agents only use their knowledge of the past and 
present situation and adapt their {behavior} accordingly. The fact that the outcome of these two hypotheses on the stability of the economy are so radically 
different should be a strong caveat. 
Which of these two extreme visions of the world is closest to reality is of course a matter of empirical investigation (on this point, see~\cite{Hommes-DSGE}, 
and for recent attempts to include deviations from perfect rationality in DSGE models, see e.g.~\cite{Woodford}).

\section{Summary, Conclusion}

In this paper, we parsimoniously extended the stylized macroeconomic Agent-Based Mark-0 model introduced in~\cite{Tipping}, with the 
aim of investigating the role and efficacy of monetary policy. We focused on three effects induced by a non-zero 
interest rate in the model, that are arguably the most important transmission channels of the Central Bank policy:
i) a change of the accounting rules to factor in the cost of debt and the extra revenue of deposits; 
ii) a change of the consumption {behavior} of household that depends negatively on the real interest rate and iii) a change in the 
hiring/firing and wage policies of firms, that avoid running into debt when interest rates increase. This amounts to adding only
two new parameters to the baseline Mark-0 model of~\cite{Tipping}. 

We first studied the model in the absence of monetary policy, i.e. without a ``Taylor-rule'' that creates a feedback between 
inflation, unemployment and interest rates. The introduction of a coupling between the financial fragility of firms and the  hiring/firing and 
wage policies has two main effects: a) the first order (discontinuous) phase transition between a `good' and a `bad' phase of the 
economy, discussed in~\cite{Tipping} is replaced by a second order (continuous) transition; b) a new first order transition to a phase with 
high residual unemployment (RU) appears when the interest rate is larger than some critical value, even in the region where full employment (FE) is achieved for 
zero-interest rate (i.e. $\Theta\gg 1$ in Fig.~\ref{fig:PD_wages}). The larger the sensitivity to interest rates, the smaller the value of 
the baseline rate beyond which the economy is destabilized.  In that region, allowing firms to accumulate more debt does not help {stabilizing} the economy. 

We then allowed the Central Bank to adjust the baseline rate so as to steer the economy towards prescribed levels of inflation and employment. 
Our major finding is that provided its policy is not too aggressive (i.e. when the targets are not too far from the `natural' state of the 
economy, and for a low enough adjustment speed) the Central Bank is successful in achieving its goals. 
However, the mere presence of different states of the economy separated by phase boundaries, besides being interesting \emph{per se}, 
can cause the monetary policy itself to trigger instabilities and be counter-productive. The destabilizing influence of the Central Bank
also depends on the firms/households sensitivities to rates. Perhaps ironically, too small sensitivities make the Central Bank policy 
inefficient, but too large sensitivities make the same policy dangerous. Seen differently, the
Central Bank must navigate in a narrow window: too little is not enough, too much leads to instabilities and wildly oscillating economies~\cite{Felix}.
As mentioned in the last paragraph, this conclusion strongly contrasts with the prediction of DSGE models. Interestingly, we also find
that a Central Bank with a dual mandate (inflation and output) enjoys a wider region of stability compared to a Central bank with inflation as the only mandate.

As we emphasized in the introduction, the key message of both our previous paper~\cite{Tipping} and the present one is that even over-simplified macroeconomic ABMs
generically display a rich phenomenology with an economy {characterized} by different states separated by phase boundaries across which radical changes of the emergent {behavior} take place.
These are, we believe, the ``dark corners'' alluded to by O. Blanchard in~\cite{Blanchard}, that both academics and policy makers should account for and wrestle with. 
We believe that the major advantage of ABMs over DSGE-like models is the very possibility of crises at the aggregate level, mediated by generic 
feedback mechanisms whose {destabilizing} role may not be immediately obvious or intuitive. Rather than the precisely calibrated predictive tools that standard equilibrium models claim to provide~\cite{Caballero}, 
ABMs offer extremely valuable qualitative tools for generating scenarios, that can be used to foresee the unintended consequences of some policy decisions. Some of this 
outcomes, which would be ``Black Swans''~\cite{Taleb} in a DSGE framework, can in fact be fully anticipated by schematic ABMs \cite{Foley}. As expressed with remarkable insight 
by Mark Buchanan~\cite{Buchanan}: {\it Done properly, computer simulation represents a kind of ``telescope for the mind,'' multiplying human powers of analysis and insight just as a
telescope does our powers of vision. With simulations, we can discover relationships that the unaided human mind, or even 
the human mind aided with the best mathematical analysis, would never grasp.}  

\section*{Acknowledgements} 
This work was partially financed by the CRISIS project. We want to thank all the members of CRISIS for 
most useful discussions, in particular during the CRISIS meetings. The input and comments of T. Assenza, J. Batista, E. Beinhocker, 
D. Challet, D. Delli Gatti, D. Farmer, J. Grazzini, C. Hommes, F. Lillo, G. Tedeschi; S. Battiston, A. Kirman, A. Mandel, M. Marsili and 
A. Roventini are warmly acknowledged. JPB  wants to sincerely thank J.-C. Trichet and A. Haldane for very encouraging comments on this endeavor. 
Finally, we thank our two referees for insightful and constructive remarks that helped improving the manuscript.

\begin{appendix}
\section{Price, production and wage updates} 
\subsection{The households timeline}

\begin{itemize}
 \item At the beginning of the time step households are {characterized} by a certain amount of savings $S(t)\geq 0$.
 \item After each firm {chooses} its production $Y_i(t)$, price $P_i(t)$ and wage $W_i(t)$ for the current time step (see later) 
 wages are paid. Since firms use a one-to-one linear technology taking only labor as input and productivity is set to $1$, the production
 equal the workforce of the firm. Hence, the total amount of wages paid is given by
 \be
 W_T(t) = \sum_i W_i(t)Y_i(t)
 \ee 
 \item Once the total payroll of the economy is determined, interests on deposits are paid and households set a consumption budget as
 \be
C_B(t) = c(t) [ S(t) + W_T(t) + \rho^{\text{d}}(t)S(t)]
\ee
where $c(t)\in[0,1]$ is the propensity to consume and may depend on inflation/interests on deposits, see Eq. (\ref{cons_budget}). 
\item The consumption budget is distributed among firms using an intensity of choice model~\cite{Anderson}. The demand of goods 
for firm $i$ is therefore:
\be\label{intensity} 
D_i(t) = \frac{C_B(t)}{p_i(t)} \frac{e^{- \beta p_i(t) / \overline p(t)}}{\sum_i 
e^{-\beta p_i(t) / \overline p(t)}} \ ,
\ee
where $\beta$ is the price sensitivity parameter determining an exponential dependence of households demand
{to} the price offered by the firm. Indeed, $\beta = 0$ corresponds to complete price insensitivity and $\beta \to \infty$ 
means that households select only the firm with the lowest price. In this sense, as long as $\beta>0$ firms 
compete on prices.
\item The \emph{actual} consumption $C(t)$ (limited by production) is given by
\be
\label{consumptionb}
C(t):= \sum_{i=1}^{N_{\rm F}}p_i(t)\min{\{Y_i(t),D_i(t)\}}\leq C_B(t) = \sum_{i=1}^{N_{\rm F}}p_i(t)D_i(t)
\ee
and households accounting therefore reads
\be
\label{households_accounting}
S(t+1) = [1+\rho^{\text{d}}(t)]S(t) + W_T(t) - \sum_i P_i(t)\min{\{D_i,Y_i\}} + \Delta(t)
\ee
where $\Delta(t)$ are dividends paid (see below for a definition of this last term).
\end{itemize}

\subsection{The firms timeline} 

\begin{itemize}
 \item At the beginning of the time step $t$ firms with $\Phi_i(t) \geq \Theta$ become inactive
 and are removed from the simulation. Default costs are computed as
 \be
 \DD(t) = -\sum_{i\ \mbox{\footnotesize bankrupt}}\EE_i(t) \ . 
 \ee
 Firms with $\Phi_i(t) < \Theta $ are instead allowed to continue their activity and contribute to total 
 loans $\EE^-(t)$ and total firms savings $\EE^+(t)$ as
 \bea
 \EE^- &=& -\sum_{i\ \mbox{\footnotesize not bankrupt}}\min{\{\EE_i(t),0\}} \\ 
 \EE^+ &=& \sum_{i\ \mbox{\footnotesize not bankrupt}}\max{\{\EE_i(t),0\}}.
 \eea
 \item Active firms set production, price and wage for the current time step following simple adaptive
 rules which are meant to represent an heuristic adjustment. In particular:
 \begin{enumerate}
  \item[-] \textbf{Price:}\\ 
  Prices are updated through a random multiplicative process which takes into account
  the production-demand gap experienced in the previous time step and  
  if the price offered is competitive (with respect to the average price). The update rule for prices reads:
  \beq
  \label{p_update}
  \begin{split}
    \text{If } Y_i(t) < D_i(t)  &\hskip10pt \Rightarrow \hskip10pt 
    \begin{cases}
&  \text{If } p_i(t) < \overline{p}(t) \hskip10pt \Rightarrow \hskip10pt   
p_i(t+1) = p_i(t) (1 + \g_p \xi_i(t) ) \\
&  \text{If } p_i(t) \geq \overline{p}(t) \hskip10pt \Rightarrow \hskip10pt   
p_i(t+1) = p_i(t) \\
\end{cases}
\\
  \text{If }   Y_i(t) > D_i(t)  &\hskip10pt  \Rightarrow \hskip10pt
  \begin{cases}
&  \text{If } p_i(t) > \overline{p}(t) \hskip10pt \Rightarrow \hskip10pt   
p_i(t+1) = p_i(t) (1 - \g_p \xi_i(t) ) \\
&  \text{If } p_i(t) \leq \overline{p}(t) \hskip10pt \Rightarrow \hskip10pt   
p_i(t+1) = p_i(t)\\
    \end{cases}
\end{split}
\eeq
where $\xi_i(t)$ are independent uniform $U[0,1]$ random variables and $\gamma_p$ is a parameter
setting the relative magnitude of the price adjustment (we set it to $5 \%$ unless stated otherwise).
{Fig.~\ref{fig:basics}, which plots the average profit of firms as a function of the offered price, shows that these rules lead to a reasonable 
emergent ``optimizing'' 
behavior of firms}.
As expected, the profit reaches a maximum
for prices slightly above the average price $\overline{p}(t)$. Higher prices are not competitive and firms lose clients, 
while lower prices do not cover production costs.

\begin{figure}
\centering
\includegraphics[scale=0.3]{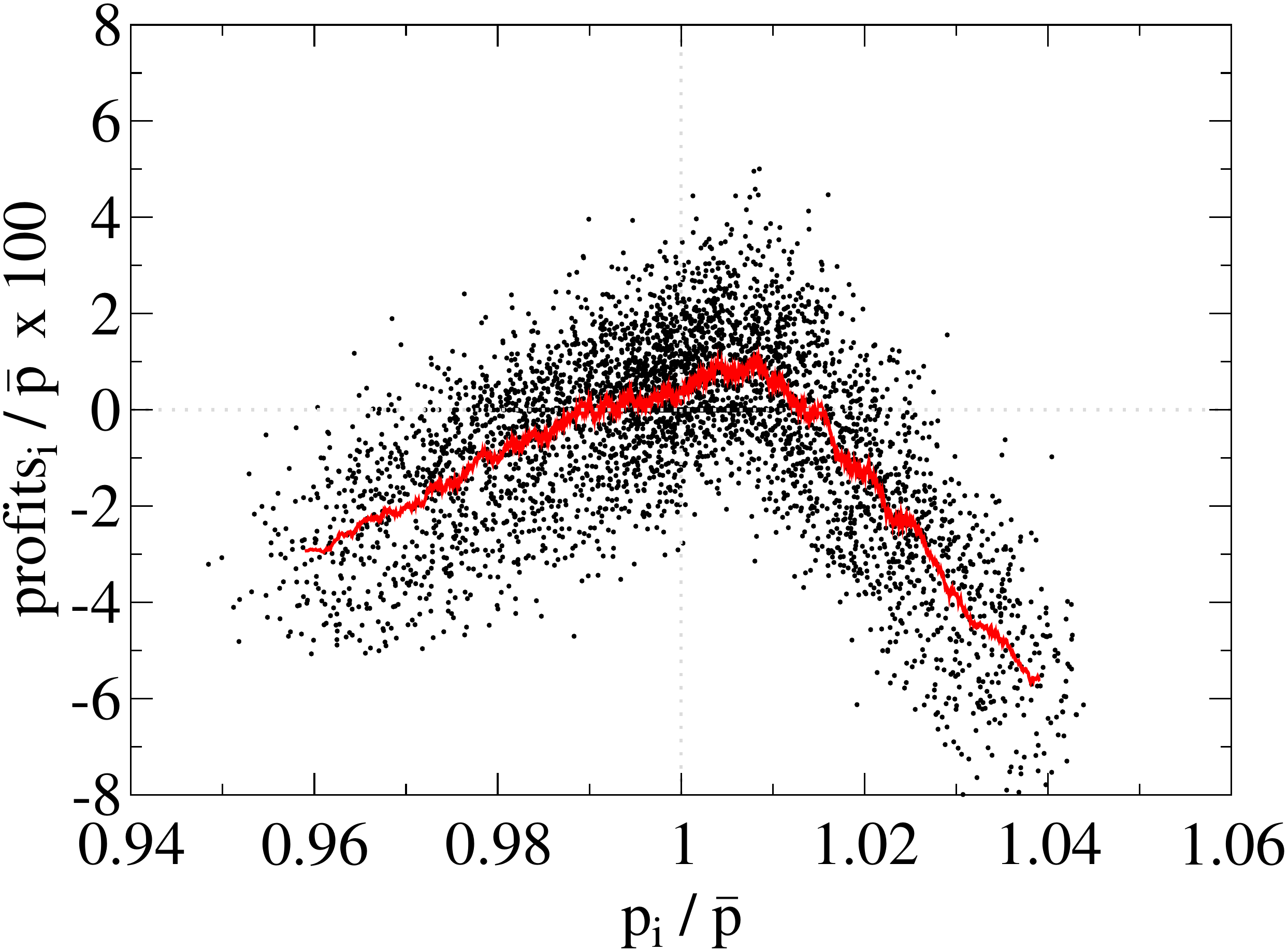}
\caption{
Scatter plot in one time step of firm profits vs the price offered. The red line
correspond to a moving flat average of $100$ consecutive points.
Parameters are:
$R=2$ (with $\eta^0_-=0.1$), $\Theta=2$, $\gamma_p=\gamma_w=0.05$, $\alpha_\Gamma=50$, $\Gamma_0=0$,
$\beta=2$, $\alpha_c=4$, $N=5000$ and $\rho^*=0.5\%$.
}
\label{fig:basics}
\end{figure}

  \item[-]\textbf{Production:}\\ 
  Independently of their price 
level, firms try to adjust their production to the observed demand. 
When firms want to hire, they open positions on the job market; we assume that 
the total number of unemployed workers, which is $N_{\rm F} u(t)$, is distributed among firms 
according to an intensity of choice model which depends on both the wage 
offered by the firm\footnote{A higher wage translates in the availability 
of a larger share of unemployed workers in the hiring process.} 
and on the same parameter $\beta$ as it is for Eq.~\eqref{intensity}; 
therefore the maximum number of available workers to each firm is:
\be
\label{wintensity} 
u^*_i(t) = \frac{e^{\beta W_i(t) / \overline w(t)}}{\sum_i e^{\beta 
W_i(t) / \overline w(t)}} N_{\rm F} u(t) \ .
\ee
The production update is then defined as:
 \beq
 \label{y_update2}
\begin{split}
    \text{If } Y_i(t) < D_i(t)  &\hskip10pt \Rightarrow \hskip10pt 
     Y_i(t+1)=Y_i(t)+ \min\{ \eta^+_i ( D_i(t)-Y_i(t)), u^*_i(t) \} \\
    \text{If }   Y_i(t) > D_i(t)  &\hskip10pt  \Rightarrow \hskip10pt
    Y_i(t+1)= Y_i(t) - \eta^-_i [Y_i(t)-D_i(t)]  \\
\end{split}
\eeq
where $\eta^\pm \in [0,1]$ are what we denote as the hiring/firing propensity 
of the firms. According to this mechanism, the change in output responds to excess
demand (there is an increase in output if excess demand is positive, a decrease
in output if excess demand is negative, i.e. if there is excess supply). The 
propensities to hire/fire $\eta_\pm$ are the sensitivity of the output change to excess 
demand/supply.

\item[-]\textbf{Wage:}\\
The wage update rule we chose follows (in spirit) the choices made for price and
production. We propose that at each time step firm $i$  updates its wage as:
\beq
\begin{split}
\label{update0++}
W^T_i(t+1)=W_i(t)[1+\gamma_w (1 - \Gamma \Phi_i) \varepsilon \xi^\prime_i(t)]
\quad\mbox{if}\quad
\begin{cases}
Y_i(t) &< D_i(t)\\
\PP_i(t) &> 0 
\end{cases}
 \\
W_i(t+1)=W_i(t)[1-\gamma_w (1 + \Gamma \Phi_i) u \xi^\prime_i(t)]
\quad\mbox{if}\quad
\begin{cases}
Y_i(t) &> D_i(t)\\
\PP_i(t) &< 0 
\end{cases}
\end{split}
\eeq
where $\gamma_w$ is a certain parameter, $\PP_i(t)$ is the profit of the firm at
time $t$ and $\xi^\prime_i(t)$ an independent $U[0,1]$ random variable.
If $W^T_i(t+1)$ is such that the profit of firm $i$ at time $t$ with this
amount of wages would have been negative, $W_i(t+1)$ is chosen to be exactly at
the equilibrium point where $\PP_i(t)=0$ otherwise $W_i(t+1) = W^T_i(t+1)$. 

The above rules are intuitive: if a firm makes a profit and it has a large demand for
its good, it will increase the pay of its workers. 
The pay rise is expected to be large if the firm is financially healthy 
and/or if unemployment is low (i.e. if $\varepsilon$ is large) because pressure on
salaries is high. 
Conversely, if the firm makes a loss and has a low demand for its good, it will reduce the wages. 
This reduction is drastic {if} the company is close to bankruptcy, and/or if
unemployment is high, because {the} pressure on salaries is then low. In all other
cases, wages are not updated.

The parameters $\gamma_{p,w}$ allow us to simulate different price/wage update timescales, i.e. the aggressivity with which 
firms react a change of their economic conditions. In the following 
we set $\gamma_p=0.05$ and $\gamma_w= \gamma_p$. 
The case $\gamma_w=0$  corresponds to removing completely the wage update rule, such that the version of 
the model with constant wage is recovered.
\end{enumerate}
\item After prices, productions and wages are set and interests paid, consumption and accounting take place. Since each firm
has total sales $p_i\min{\{Y_i,D_i\}}$ firms profits are
\be
\PP_i(t) = p_i(t)\min{\{Y_i(t),D_i(t)\}} - W_i(t)Y_i(t) + \rho^{\text{d}}\max{\{\EE_i(t),0\}} + \rho^{\ell}\min{\{\EE_i(t),0\}} \ .
\ee
When firms have both positive $\EE_i$ and $\PP_i$ dividends are paid as a fraction $\delta$ of the firm cash balance $\EE_i$.
The update rule for firms {cash} balance is therefore
\be
\EE_i(t+1) = \EE_i(t) + \PP_i(t) - \delta \EE_i(t)\theta(\PP_i(t))\theta(\EE_i(t))
\ee
where $\theta(x)=1$ if $x>0$ and $0$ otherwise.
Correspondingly, households savings are updated as
\be
S(t+1) = S(t) + \sum_iW_i(t)Y_i(t) - \sum_i p_i(t)\min{\{Y_i(t),D_i(t)\}}+ \delta \sum_i\EE_i(t)\theta(\PP_i(t))\theta(\EE_i(t)).
\ee
The dividends share $\delta$ is set to $2\%$ unless stated otherwise
and the $\Delta(t)$ term in Eq.~\eqref{households_accounting} is given by
\be
\Delta(t) = \delta \sum_i\EE_i(t)\theta(\PP_i(t))\theta(\EE_i(t))
\ee
\item Finally, an inactive firm has a finite probability $\varphi$ (which we set to $0.1$) per unit time to get
revived; when it does so its price is fixed to $p_i(t) = \overline p(t)$, its wage to
$w_i(t) = \overline w(t)$, its workforce is the available workforce $Y_i(t) = \mu u(t)$
and its cash-balance is the amount needed to pay the wage bill $\EE_i(t) = 
W_i(t) Y_i(t)$. This small 'liquidity' is provided by firms with positive $\EE_i$ in shares proportional 
to their wealth $\EE_i$.
\end{itemize}

\section{Firms' adaptive {behavior} leads to a second order phase transition}
\label{adaptation}

We start by analysing the effect of adaptation of firms.  In order to get a first insight it is useful to consider a simplified 
setting where $\G=\Gamma_0$ (i.e. $\alpha_\Gamma=0$), $\rho^{\ell}(t)=\rho_0(t)=0$, $f=1$, $c(t)=c_0=0.5$ (hence $\alpha_c=0$) 
and wages are constant and equal to $1$ ($\gamma_w=0$). In this case the basic model described in~\cite{Tipping} 
(with constant wages) is recovered. 

Intuitively, the coupling between financial fragility and hiring/firing propensity should have a {stabilizing} effect on the 
economy. Moreover, the full unemployment phase at $R< R_c$ is deeply affected by the presence of $\alpha_\Gamma$: 
for $\alpha_\Gamma\neq 0$ the unemployment rate in this phase is no longer one, but becomes smaller than one and continuously changing with $R$.
In order to derive an estimate of these continuous values we use an intuitive argument (at $\Th=\io$) which is justified a posteriori 
by the good match with numerical results.
Given the critical ratio $R = \eta^0_+/\eta^0_- = R_c$ separating the high/low
unemployment phases when there is no adaptation (i.e. $\Gamma_0=0$) one can expect
that equilibrium values of the unemployment rate different from $0$ and $1$ can
only be stable if $\eta_+^{i}/\eta_{-}^{i}$ remains around the critical 
value $R_c$ at $\Gamma_0=0$. 
Near criticality therefore we enforce that:
\be
\frac{\eta_+^i}{\eta_-^i} = \frac{\eta^0_+ (1 - \Gamma_0 \Phi_i)}{\eta^0_- (1 + \Gamma_0 \Phi_i) } = R_c
\hskip10pt
\Rightarrow
\hskip10pt
-\Gamma_0 \Phi \approx \frac{R_c\eta_-^0-\eta_+^0}{R_c\eta_-^0+\eta_+^0}.
\ee
An explicit form of $\Phi$ in terms of the employment rate $\varepsilon={\overline Y}$ can be
obtained with the additional assumption that the system is always close to
equilibrium (i.e. $p\approx 1$ and $D \approx Y$, at least when $\eta^i_+/\eta^i_-\sim R_c$),
which allows one to express households savings in terms of the firms' production. Indeed (see the discussion in~\cite{Tipping}) at equilibrium $W = B = N_{\rm F} Y = c(W + S)$, 
from which it follows that $N_{\rm F} Y = W = S c/(1-c)$.
For $c=0.5$ as in our simulations one thus has $S=N_{\rm F} Y$. Since the total amount of money 
is conserved (in our simulations $N_{\rm F} \overline{\EE} + S = N_{\rm F}\overline{\EE} + N_{\rm F}Y= N_{\rm F}$, 
see Appendix~\ref{app:Mark0}) one finally obtains
that $\overline{\LL} = 1 - Y$ and $\Phi = (Y-1)/Y = (\varepsilon-1)/\varepsilon$, hence
\be
\frac{\Gamma_0}{\varepsilon} = 
 \frac{R_c\eta_-^0-\eta_+^0}{R_c\eta_-^0+\eta_+^0} + \Gamma_0
 = 
 \frac{R_c- R}{R_c+ R} + \Gamma_0
  \ .
\label{EmpRes}
\ee
Note that according to this formula the employment goes to $\varepsilon=1$ at the critical point $R = R_c$.
Above this value, the economy is in the ``good'' state and employment sticks to $\varepsilon=1$ (this is because in the argument the effect of 
$\Theta$ has been neglected).
Moreover, when $R = R_c$, $\varepsilon$ is proportional to $\G_0$ and therefore in the limit $\Gamma_0\to 0$ one has
$\varepsilon=0$ for  all $R < R_c$. This is the ``bad'' phase of full unemployment at $\Gamma_0=0$, which becomes
in this case a phase where employment grows steadily but remains of order $\G_0$ except very close to the critical point.

\begin{figure}
\centering
\includegraphics[scale=0.4]{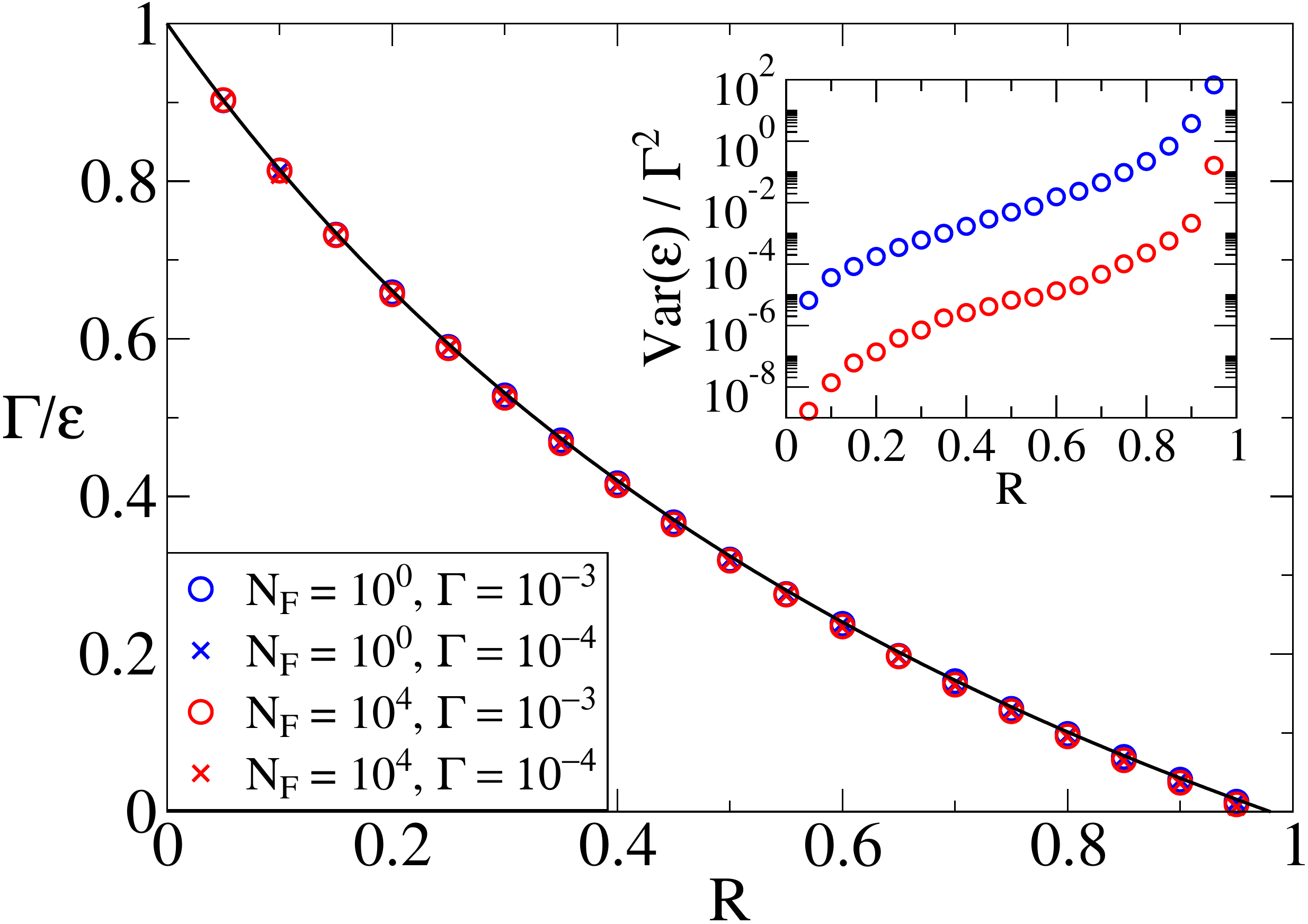}
\caption{
Inverse of the average employment rate $\Gamma_0/\overline{\varepsilon}$ as a function of the ratio $R=\eta^0_+/\eta^0_-$ with
$\eta^0_-=0.1$ and $\gamma=0.01$ when $\Gamma_0>0$.  When the employment rate is rescaled with the parameter $\Gamma_0$ (here
$\Gamma_0=10^{-3},\ 10^{-4}$) the different lines collapse and Eq.~\eqref{EmpRes} agrees with numerical simulations. In the inset we also plot the rescaled
variance, still as a function of $\eta^0_+$. Approaching the critical point the variance of the unemployment fluctuations diverges, together with their 
relaxation time going to infinity. The other parameters are: $\delta=0.02$, $\Theta=5$, $\gamma_w=0$, $c=0.5$, $\beta=0$ and $\varphi=0.1$ 
}
\label{fig:EmpRes}
\end{figure}

\begin{figure}
\centering
\includegraphics[scale=0.25]{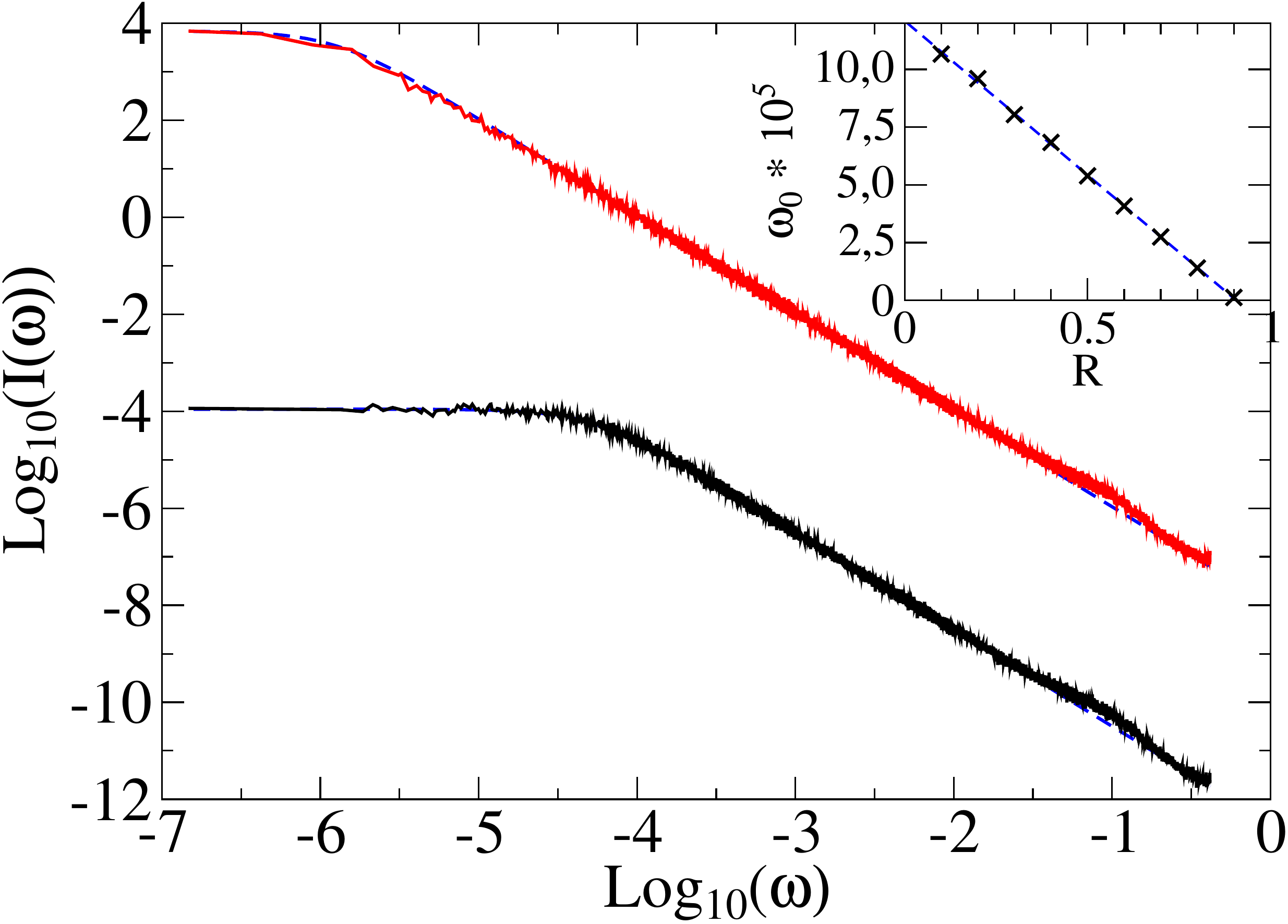}
\includegraphics[scale=0.25]{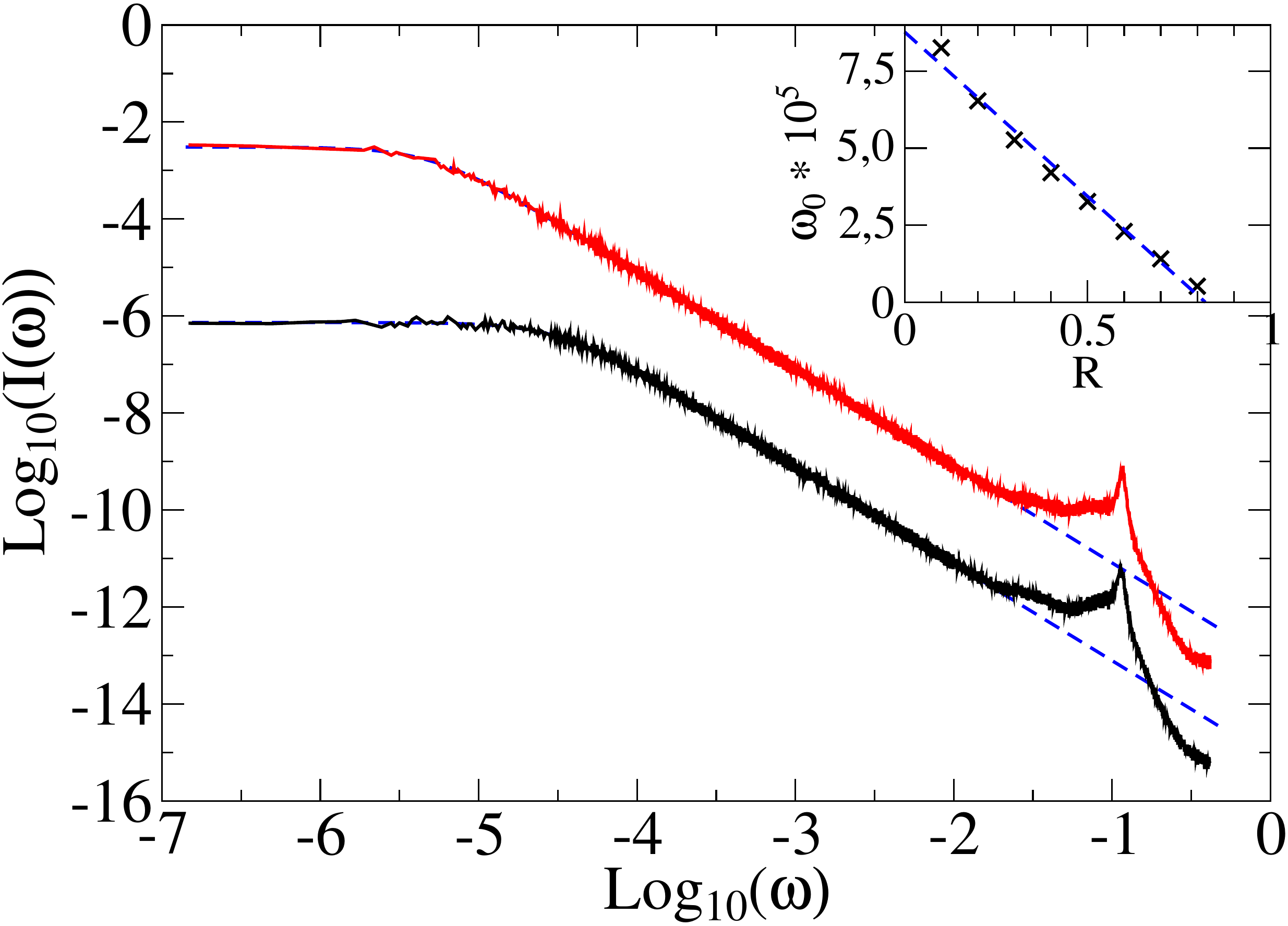}
\caption{Logarithm of the {normalized} power spectrum for Mark 0 with adaptive firms ($\Gamma_0=10^{-3}$), $\gamma_p=0.05$ and $N_{\rm F}=1$
(left) and $N_{\rm F}=1000$ (right).  The other parameters are set as in Fig.~\ref{fig:EmpRes}.
The main plot {shows} two examples of the spectrum for $\eta_-^0=0.1$ and $\eta_+^0=0.05$ (black line) and $\eta^0_+=0.09$ (red line) in the left plot,
$\eta_+^0=0.05$ (black line) and $\eta^0_+=0.08$ (red line) in the right plot. 
The time series is made of $2^{28}$ time steps after $T_{eq}=500\, 000$ and the logarithm of the spectrum is averaged over a moving 
window of $100$ points for a better {visualization}. With both system sizes the fit with Eq.~\eqref{spectrum_fit} (blue dashed lines) is good 
with the only difference that when $N_{\rm F}>1$ a clear oscillatory {pattern appears} at high frequencies, that becomes sharper and sharper as $N_{\rm F}$ increases. 
In the inset of each figure we plot the value of $\omega_0$ in Eq.~\eqref{spectrum_fit} obtained from the fit as a function of the ratio $R=\eta_+^0/\eta_-^0$. 
In both cases $\omega_0$ goes linearly to $0$ as the critical value is approached. }
\label{fig:FT}
\end{figure}

Eq.~\eqref{EmpRes} is plotted in Fig.~\ref{fig:EmpRes} together with numerical results.
Note that in this case the representative firm approximation ($N_{\rm F}=1$) is in good agreement with numerical
results also for $N_{\rm F}=10,000$, as it was for the discontinuous transition obtained for $\G=0$. 
In the inset of Fig.~\ref{fig:EmpRes} one can see that the variance of the fluctuations of 
employment rate is diverging as long as the critical value of $R$ is
approached. This is confirmed by a spectral analysis of the unemployment time series (see
Fig.~\ref{fig:FT}). In order to obtain the power spectrum we apply the GSL Fast Fourier Transform
algorithm to the time series $\varepsilon(t)-\avg{\varepsilon}$.
As one can see in Fig.~\ref{fig:FT} the power spectrum is well approximated by an Ornstein-Uhlenbeck form:
\be
I(\omega)=I_0\frac{ \omega_0^2}{\omega_0^2+\omega^2}
\label{spectrum_fit}
\ee
with $\omega_0$ going linearly to $0$ when $\eta^0_+$ approaches its critical
value, meaning that the relaxation time $\omega_0^{-1}$ diverges as one approaches the
critical point. Note that this is not the case for the Mark 0 model with $\Gamma_0=0$
which instead has a white noise power spectrum even in proximity of the
transition line. The first order (discontinuous) transition for $\Gamma_0=0, \Theta=\infty$ is thus
replaced by a second order (continuous) transition when the firms adapt their {behavior}
as a function of their financial fragility.

Finally, note that the presence of a continuum of states for the unemployment rate whenever $\Gamma_0>0$ and $R<R_c$ 
holds also with $\gamma_w>0$ (when wages are not constant). It was however simpler to perform analytical computations
with constant wages.

\section{Pseudo-code of Mark 0}
\label{app:Mark0}

We present here the pseudo-code for the Mark 0 code described in Sec.~\ref{m0extended} and Appendix A. The source code of the baseline Mark-0 
is available on demand.

\begin{algorithm}[h]
\caption{Mark 0}         
\label{alg:Mark0}                          
\begin{algorithmic} 
\Require$N_{\rm F}$ Number of firms; $c_0,\beta,\gamma_p,\gamma_w,\eta^0_+,\eta^0_-,\delta,\Theta,\varphi,f,\alpha_c,\phi_\pi,\phi_\varepsilon,\alpha_\Gamma,\Gamma_0,\varepsilon^*,\pi^*,\rho^*,\omega$; 
$T$ total evolution time; 
\State
\Comment{ {\bf\color{red} Initialization}}
\For {($i \leftarrow 0; i< N_{\rm F} ;i\leftarrow i+1$)}
\State $W[i] \leftarrow 1$
\State $p[i] \leftarrow 1+0.2 (\text{\tt random} - 0.5) $
\State $Y[i] \leftarrow [1 +0.2 (\text{\tt random} - 0.5)] /2$
\State $D[i] \leftarrow 0.5$
\Comment{Initial employment is $0.5$}
\State $\EE[i] \leftarrow W[i] Y[i] \, \text{\tt random}$
\State $\PP[i] \leftarrow p[i] \min(D[i],Y[i]) - W[i] Y[i]$
\State $a[i] \leftarrow 1$
\Comment{binary variable: active ($1$) / inactive ($0$) firm}
\EndFor
\State $S\leftarrow N_{\rm F} - \sum_i \EE[i]$
\Comment{ {\bf\color{red} Main loop}}
\For {($t \leftarrow 1; t \leq T;t\leftarrow t+1$)}  
\State $\varepsilon\leftarrow \frac1{ N_{\rm F}} \sum_i Y[i]$
\State $u \leftarrow 1-\varepsilon$
\State $\overline p \leftarrow \frac{\sum_i p[i] Y[i]}{\sum_i Y[i]}$ 
\State $\overline w \leftarrow \frac{\sum_i W[i] Y[i]}{\sum_i Y[i]}$ 
\State $u^*[i] \leftarrow \frac{\exp(\b W[i]/\overline w)}{\sum_i a[i]\exp(\b W[i]/\overline w)}N_{\rm F}u $
\State $\widetilde{x} \leftarrow \omega x +(1-\omega)\widetilde{x}$ where $x$ are $\pi,\rho^{\text{d}},\rho^{\ell},u$
\Comment{{\bf \color{blue} Central Bank policy}}
\State $\hat{\varepsilon}^*\leftarrow \min{(\varepsilon^*,1.025 \tilde{\varepsilon})}$
\State $\rho_0 \leftarrow \rho^* + 10\phi_\pi(\tilde{\pi}-\pi^*) + \phi_\varepsilon \log{(\tilde{\varepsilon}/\hat{\varepsilon}^*)}$
\State $\rho_0 \leftarrow \max{(\rho_0,0)}$
\State $\G \leftarrow \max{\{\alpha_\Gamma(\widetilde\rho^{\ell} - \widetilde{\pi}),\Gamma_0 \}}$
\State $\mathcal{D} \leftarrow \EE^- \leftarrow \EE^+ \leftarrow 0$
\Comment{{\bf \color{blue} Firms update prices, productions and wages}}
\For {($i \leftarrow 0; i< N_{\rm F} ;i\leftarrow i+1$)}
\If { $a[i]==1$ }
\If {$\EE[i]>-\Theta W[i] Y[i]$}
\State $\EE^+\leftarrow \EE^++\max{\{\EE[i],0\}}$
\State $\EE^-\leftarrow \EE^- -\min{\{\EE[i],0\}}$
\State $\Phi[i] \leftarrow -\frac{\EE[i]}{W[i]Y[i]}$
\State $\Phi[i] \leftarrow\min{\{\Phi[i],1/\Gamma\}}$
\State $\Phi[i] \leftarrow\max{\{\Phi[i],-1/\Gamma\}}$
\State $\h_+\leftarrow\h^0_+(1-\Gamma \Phi[i])$
\State $\h_-\leftarrow\h^0_-(1+\Gamma \Phi[i])$
\State
\If { $Y[i] < D[i]$ }
\If {$\PP[i]>0$}
\State $W[i]\leftarrow W[i][1+\g_w(1-\Gamma \Phi[i])\varepsilon$ {\tt random}$]$
\State $W[i]\leftarrow \min{\{W[i],(P[i]\min{[D[i],Y[i]]} + \rho^{\text{d}}\max{\{\EE[i],0\}}+ \rho^{\ell}\min{\{\EE[i],0\}})/Y[i]\}}$
\EndIf
\State $Y[i] \leftarrow Y[i] + \min\{ \h_+  (D[i] - Y[i] ),  u^*[i] \}$
\If{ $p[i] < \overline p$} $p[i] \leftarrow p[i] ( 1 + \g_p \,${\tt random})
\EndIf
\ElsIf{ $Y[i] > D[i]$ }
\If {$\PP[i]<0$}
\State $W[i]\leftarrow W[i][1-\g_w(1+\Gamma \Phi[i])u$ {\tt random}$]$
\EndIf
\State $Y[i] \leftarrow \max\{ 0, Y[i] - \h_- ( D[i] - Y[i] ) \}$
\If{ $p[i] < \overline p$ } $p[i] \leftarrow p[i] ( 1 - \g_p \,${\tt random})
\EndIf
\EndIf
\ElsIf {$\EE[i]\leq-\Theta W[i] Y[i]$}
\State $a[i]\leftarrow 0$
\State $\mathcal{D}\leftarrow  \mathcal{D}- \EE[i]$
\EndIf
\EndIf
\EndFor
      \algstore{Mark0}
  \end{algorithmic}
\end{algorithm}
\begin{algorithm}
  \caption{Mark0 (continued)}
  \begin{algorithmic}
      \algrestore{Mark0}
\State
\State $u \leftarrow 1-\frac1{ N_{\rm F}} \sum_i Y[i]$
\Comment{Update $u$ and $\overline p$}
\State $\overline p \leftarrow \frac{\sum_i p[i] Y[i]}{\sum_i Y[i]}$
\State
\Comment{{\bf \color{blue} Private bank sets interest rates}}
\State $\rho^{\ell} = \rho_0 + (1-f)\mathcal{D}/\EE^-$
\State $\rho^{\text{d}} = \frac{\rho^{\ell}\EE^--\mathcal{D}}{S+\EE^+}$
\State
\Comment{{\bf \color{blue} Households decide the demand}}
\State $S \leftarrow (1+\rho^{\text{d}})S + \sum_i W[i]Y[i]$
\State $c  \leftarrow c_0 [1+\alpha_c(\widetilde{\pi}-\widetilde{\rho}^{\text{d}})/\gamma_\pi]$
\State $C_B \leftarrow c S$
\State
\For {($i \leftarrow 0; i< N_{\rm F} ;i\leftarrow i+1$)}
\State $D[i] \leftarrow \frac{C_B a[i] \exp(-\b p[i]/\overline p)}{p[i] \sum_i a[i]\exp(-\b p[i]/\overline p)} $
\Comment{Inactive firms have no demand}
\EndFor
\State
\Comment{{\bf \color{blue} Accounting}}
\State $\EE^+ \leftarrow 0$
\For {($i \leftarrow 0; i< N_{\rm F} ;i\leftarrow i+1$)}
\If { $a[i]==1$ }
\State $S \leftarrow S - p[i] \min\{ Y[i], D[i] \}$
\State $\PP[i] \leftarrow p[i] \min\{ Y[i], D[i] \} - W[i] Y[i]+ \rho^{\text{d}}\max{\{\EE[i],0\}}+ \rho^{\ell}\min{\{\EE[i],0\}}$
\State $\EE[i] \leftarrow \EE[i] + \PP[i]$
\If { $\PP[i] > 0$ \&\& $\EE[i]>0$ }
\Comment{Pay dividends}
\State $S \leftarrow S + \d \, \EE[i]$
\State $\EE[i] \leftarrow \EE[i] - \d \, \EE[i]$
\EndIf
\State $\EE^+\leftarrow \EE^++\max{\{\EE[i],0\}}$
\EndIf
\EndFor
\State
\Comment{{\bf \color{blue} Revivals}}
\State $\mathcal{R} \leftarrow 0$
\For {($i \leftarrow 0; i< N_{\rm F} ;i\leftarrow i+1$)}
\If { $a[i]==0$}
\If { {\tt random }$<\varphi$ }
\State $Y[i] \leftarrow u\mbox{ random}$
\State $a[i] \leftarrow 1$
\State $P[i] \leftarrow \overline p$
\State $W[i] \leftarrow \overline w$
\State $\EE[i] \leftarrow W[i]Y[i]$
\State $\mathcal{R}\leftarrow \mathcal{R}+\EE[i]$
\State $\EE^+\leftarrow \EE^++\max{\{\EE[i],0\}}$
\EndIf
\EndIf
\EndFor
\For {($i \leftarrow 0; i< N_{\rm F} ;i\leftarrow i+1$)}
\If{$a[i]==1$}
\If {$\EE[i]>0$}
\State $\EE[i] \leftarrow \EE[i] - \mathcal{R} \EE[i]/\EE^+$
\EndIf
\EndIf 
\EndFor
\EndFor  

\end{algorithmic}
\end{algorithm}

\end{appendix}

\end{document}